\documentclass[a4paper]{article}
\usepackage[american]{babel}
\usepackage{amssymb,amsfonts,amsmath,amsthm}
\usepackage{a4wide,graphicx,color}
\usepackage{paralist}

\newcommand{\abs}[1]{\left\vert #1\right\vert}
\newcommand{\cC}{\mathcal{C}}
\newcommand{\Cc}{C^{\infty}_{c}(\Rn)}
\newcommand{\CTM}{C([0,\,T];\,\bM)}
\newcommand{\bW}{\mathbf{W}}

\newcommand{\ds}{\displaystyle}
\newcommand{\duality}[2]{\left\langle#1,\,#2\right\rangle}
\newcommand{\ie}{i.e.}
\newcommand{\Leb}[1]{\mathcal{L}^{#1}}
\newcommand{\Lip}[1]{\operatorname{Lip}(#1)}
\newcommand{\Lipone}{\textrm{Lip}_1(\Rn)}
\newcommand{\bM}{\mathbf{M}^{N^1,N^2}}
\newcommand{\cM}{\mathcal{M}}
\newcommand{\p}{\mathsf{p}}
\newcommand{\PP}{\mathsf{P}}
\newcommand{\q}{\mathsf{q}}
\newcommand{\rr}{\mathbf{r}}
\newcommand{\R}{\mathbb{R}}
\newcommand{\Rn}{\R^n}
\newcommand{\unit}[1]{~\mathrm{#1}}
\newcommand{\vchem}[1]{v^{#1}_{\textup{chem}}}
\newcommand{\vint}[1]{v^{#1}_{\textup{int}}}

\let\div\relax
\DeclareMathOperator{\div}{div}
\DeclareMathOperator{\diam}{diam}

\theoremstyle{remark}\newtheorem*{remark}{Remark}
\theoremstyle{plain}
	\newtheorem{proposition}{Proposition}[subsection]
	\newtheorem{theorem}[proposition]{Theorem}

\graphicspath{{./}{pictures/}}

\title{Differentiated cell behavior: a multiscale approach using measure theory}
\author{Annachiara Colombi,
		Marco Scianna\thanks{This author has been funded by a post-doctoral research scholarship awarded by the National Institute for Advanced Mathematics
		``F. Severi'' (INdAM, Italy).} \\
		{\small\it Department of Mathematical Sciences} \\[-1mm]
		{\small\it Politecnico di Torino} \\[-1mm]
		{\small\it Corso Duca degli Abruzzi 24, 10129 Torino, Italy} \\[5mm]
		Andrea Tosin \\
		{\small\it Istituto per le Applicazioni del Calcolo ``M. Picone''} \\[-1mm]
		{\small\it Consiglio Nazionale delle Ricerche} \\[-1mm]
		{\small\it Via dei Taurini 19, 00185 Rome, Italy}
	   }
\date{}
	
\begin{document}

\maketitle

\begin{abstract}
This paper deals with the derivation of a collective model of cell populations out of an individual-based description of the underlying physical particle system. By looking at the spatial distribution of cells in terms of \emph{time-evolving measures}, rather than at individual cell paths, we obtain an ensemble representation stemming from the phenomenological behavior of the single component cells. In particular, as a key advantage of our approach, the scale of representation of the system, \ie, microscopic/discrete vs. macroscopic/continuous, can be chosen \emph{a posteriori} according only to the spatial structure given to the aforesaid measures. The paper focuses in particular on the use of different scales based on the specific functions performed by cells. A two-population hybrid system is considered, where cells with a specialized/differentiated phenotype are treated as a discrete population of point masses while unspecialized/undifferentiated cell aggregates are represented with a continuous approximation. Numerical simulations and analytical investigations emphasize the role of some biologically relevant parameters in determining the specific evolution of such a hybrid cell system.

\medskip


\medskip

\end{abstract}

\section{Introduction}
Living systems are characterized by \emph{collective} dynamics that arise from \emph{individual} behaviors and interactions through \emph{multiscale} interconnected processes. More specifically, the evolution of macroscopic cell aggregates, visible by looking at the totality of cells as a whole, results from the phenomenology of single microscopic cells, that are able to actively sense and interact with the surrounding environment. It appears therefore natural that more and more importance is being paid in Mathematical Biology to the development of multiscale modeling structures capable of ensemble representations while still linking collective dynamics to the small scale cell behaviors (see, for instance,~\cite{Anderson2007,Cristini2010,Deutsch2005,Scianna2013a} and references therein).

With this broad idea in mind, and inspired by the methods developed for crowd dynamics in~\cite{Cristiani2011,Cristiani2014,Piccoli2009,Piccoli2011}, we here propose a measure-theoretic approach to the modeling of cell populations, which permits a coherent passage from an individual description of single cell behaviors to a collective representation of the aggregate distribution in terms of time-evolving measures. More specifically, we admit that cell positions are microscopic variables, whose evolution in time is determined by hybrid differential equations implementing interactions between \emph{individual} cells and the \emph{collective} distribution of the surrounding mass. Then we resort to some ideas from the transport of measures, in a mass-conservative framework, for abstracting the time evolution of such variables into the spatiotemporal evolution of a measure $\mu_t$ representing their aggregate distribution is space. Mass conservation is here motivated by the fact that, at the individual cell level, we consider only conservative interactions, \ie, interactions which do not change the number of involved particles. Thus, it makes sense to assume that the aforesaid measure is simply moved and redistributed by some velocity field, which has to be properly linked to the microscopic cell behavior and, besides the already mentioned cell interactions, may also include a component determined by an extracellular chemical substance (\ie, via chemotactic or haptotactic mechanisms).

With this procedure, we naturally pass from a Lagrangian to an Eulerian description, while still preserving the finiteness of both the total number of cells (\ie, we do not take the limit of an infinite number of cells) and the individual mass of the latter (which does not become infinitesimal). Therefore all quantities, such as the cell sensing radii and the intercellular interaction strengths, can still be directly referred to the characteristic cell scale. The resulting mathematical structure enables us to derive discrete or continuous models from the very same abstract framework only by acting on the spatial structure of the measure $\mu_t$ (atomic or absolutely continuous with respect to Lebesgue, respectively). It is worth pointing out that, in our case, the continuous and discrete descriptions should not be understood as a sort of approximation of one another. In fact, as implied above, the ``microscopic vs. macroscopic'' dichotomy is here not meant in terms of space averages of an infinite number of massless (in the limit) particles. This makes the difference between the proposed approach and classical approaches presented in the literature, which typically rely on the idea that the continuous description emerges in the limit from a vanishing discrete description, and is thus a good approximation of the latter for very large numbers of particles. For instance, in~\cite{Stevens2000} a macroscopic chemotaxis system describing aggregation phenomena due to cell chemical responses to a diffusing substance is derived as the limit of an interacting stochastic many-particle system, where the interaction among the particles is suitably rescaled as the population size tends to infinity. In~\cite{Carrillo2010} the authors employ a \emph{mean-field limit} to pass from microscopic/Cucker-Smale-type models of swarming to a statistical representation by means of kinetic equations, still in the limit of an infinite number $N$ of particles whose mass is scaled as $1/N$. In~\cite{Capasso2009SAA,Morale2005} a heuristic \emph{law of large numbers} is instead used to derive a deterministic macroscopic/Eulerian model for the evolution of systems of interacting particles by scaling up from a microscopic/Langrangian description of individual dynamics, determined by sets of It\^{o}-type stochastic differential equations. This approach is employed by the same authors to study tumor-induced angiogenic processes~\cite{Capasso2009JMB}. Finally, in~\cite{Drasdo2005} a cellular automaton model is the starting point to construct a multivariate master equation, from which a continuous model is derived by a systematic \emph{coarse-graining} procedure. This method consists in expressing the rules of the cellular automaton in terms of chemical reactions and in focusing on the evolution of spatial compartments which contain a sufficiently high number of massless cells.

Our approach is of particular relevance in biological systems composed by several functional subsystems, for instance multiple cell populations or multiple clones of the same population, which individually require a distinct type of mathematical description according to their specific functions. For instance, few specialized/differentiated cells can be properly described by a concentrated (\ie, discrete) representation. On the other hand, nonspecialized/undifferentiated cells can be represented as a continuously distributed group with low individual detail. Such a discrete-continuous coupling is expected to provide a useful approach for a wide range of biological phenomena. It is, for example, the case of blood vessel formation and development, where a small number of tip cells behave as a migratory guidance for the collective patterning of the rest of stalk individuals. This process is determined by the so-called \emph{tip cell selection and lateral inhibition} and mediated by selected VEGF-induced delta-notch signaling pathways. Similarly, during repair of skin after injury, epidermal monolayers invade the wound region moving across two-dimensional ECM substrates. In particular, the cells located at the front of the population are characterized by actin-rich lamellipodia and pseudopodia, which generate traction forces on the collagenous matrix, and are able to synthesize a new basement membrane, whereas the movement of following individuals is only due to cell-cell and cell-matrix adhesive interactions. Typical phenotypic differences are also observed in pathological situations, such as in tumor growth. Nutrient gradients form in fact within a solid cancer mass, causing a well-localized differentiation of malignant cells, which typically differentiate in an outer viable rim of highly metabolic and proliferative individuals and a central quiescent (possibly necrotic) core. In particular, the external cells, which are difficult to be clinically detected, have the greatest potential to metastasize, displaying an evident ability to evade destruction by the immune system, enter the host bloodstream or lymphatics, extravasate at a distant site, and establish secondary colonies with devastating consequences for the wellbeing of the patient. According to these considerations, a discrete description is biologically appropriate for the cells at the external ring of a tumor, whereas a continuous approximation is natural to represent the inner core of a malignant mass. Analogously, our approach can easily deal with the \emph{epithelial-to-mesenchymal transition}, typical of different stages of morphogenesis and of development of carcinomas (\ie, tumor of epithelial origin, such as lung, breast, ovarian, and prostate malignancies). This process, caused by a dramatic downregulation of cell-cell adhesive junctions, consist in fact in the delamination (and in the subsequent individual invasion of the mesenchyme also via proteolytic activity) of groups of aggressive cells from epithelial layers.

After the above general presentation, the contents of the paper can now be outlined more precisely. In Section~\ref{sec:math_model} we introduce the mathematical structures based on time-evolving measures and we derive the Eulerian model for the evolution of cell aggregates out of the Lagrangian description of single cell behavior. In Section~\ref{sec:two_pop} we focus on a two-population system, where few activated/discrete cells drive the evolution of an undifferentiated/continuous mass. In Section~\ref{sec:simulations} we perform a computational analysis of our method by means of exploratory case studies, which investigate the joint influence of the spatial representation and of some biologically relevant model parameters on the resulting cell dynamics. In Section~\ref{sec:comparison} we propose an analytical study, whose purpose is to inquire into the similarities and differences in the dynamics of the two-population system caused by switching from the continuous to a conceivable discrete representation of the population of undifferentiated cells. Finally, in Section~\ref{sec:discussion} we discuss the results of the paper and we briefly sketch research perspectives.

\section{Mathematical model}
\label{sec:math_model}
We consider a biological system formed by $\PP\geq 1$ functional subsystems, viz. cell populations, each of them with a finite and constant number of individuals $N^\p$ (\ie, we are assuming that cells do not undergo mitosis or apoptosis). The evolution of a generic representative cell of population $\p$, hereafter called \emph{test cell}, is then described at the suitable \emph{individual} scale by means of a variable\footnote{Throughout the paper we will use the subindex $t$ to denote dependence on time of measures or measure-related quantities. In no case will this notation stand for the time derivative, which will be indicated by an explicit symbol such as e.g., $\partial_t$.} $X^\p_t$, $t\geq 0$ being the time, which takes values in $\Rn$. If an initial position $x\in\Rn$ is assigned, the mapping $t\mapsto X^\p_t(x)$ represents the trajectory of the cell placed in $x$ at the initial time $t=0$, \ie, we are taking a Lagrangian point of view. A practical assumption for reducing the complexity of the problem is that cells within the same population are \emph{indistinguishable} from one another, in the sense that if initially two cells belonging to the $\p$-th population switch their positions the dynamics observed at future times are the same.

The total mass of the $\p$-th cell population is instead described by a Radon positive measure $\mu^\p_t$, that we assume to be defined on the Borel $\sigma$-algebra $\mathcal{B}(\Rn)$. For any measurable set $E\in\mathcal{B}(\Rn)$, the value $\mu^\p_t(E)\geq 0$ measures the mass of cells of the $\p$-th subsystem located in $E$ at time $t$. In particular, the $\sigma$-additivity of $\mu^\p_t$ directly translates the concept of the additivity of physical masses so that
$$ \mu^\p_t(\Rn)=N^\p, \quad \forall\,t\geq 0,\ \p=1,\,\dots,\,\PP. $$
Summarizing, $X^\p_t$ is characteristic of an \emph{individual} description, whereas the measure $\mu^\p_t$ looks at the system from a \emph{collective} point of view, not necessarily focused on single cells.

\subsection{Particle models}
\label{part_model}
To approach the dynamics of the test cell, we recall that biological individuals are not passively prone to Newtonian laws of inertia, but are able to actively develop behavioral strategies which depend on both specific internal stimuli and the interaction with the external environment (other cells, chemical fields or extracellular matrix components). Furthermore, in extremely viscous regimes, such as the biological environments themselves, the velocity of moving individuals and not their acceleration is typically proportional to the sensed forces (\emph{overdamped force-velocity response}, characteristic of several Individual-Based Models~\cite{Drasdo2005,Drasdo2005a,Scianna2012}).

These concepts, also developed for other systems of living entities, such as human crowds~\cite{Cristiani2014}, can be translated into mathematical terms by a direct modeling of the velocity of the cells. In particular, we here assume that the dynamics of a cell in an aggregate result from:
\begin{inparaenum}[i)]
\item a specific directional movement;
\item its interactions with the ensemble distribution of surrounding individuals.
\end{inparaenum}
A prototype for the evolution equation of the variable $X^\p_t$ then writes as:
\begin{align}
	\begin{aligned}[b]
	    \dot{X}^\p_t &= \vchem{\p}(t,\,X^\p_t)+\vint{\p}(X^p_t) \\
    		&= \vchem{\p}(t,\,X^\p_t)+\sum_{\q=1}^{\PP}\int_{\Rn\setminus\{X^\p_t\}}H^{\p\q}(y-X^\p_t)d\mu^\q_t(y).
    	\end{aligned}
    \label{eq:test_cell}
\end{align}
Notice that~\eqref{eq:test_cell} is a \emph{hybrid} differential equation, because it implements interactions between the individual cell $X^\p_t$ and the collective distributions $\mu^\q_t$ of the surrounding masses.

The term $\vchem{\p}:\Rn\to\Rn$ describes a motion field generated by either the topology of the extracellular environment (\ie, via durotactic or haptotactic mechanisms) or the spatial distribution of some diffusive chemical that test cells are sensitive to (\ie, via chemotaxis).

The term $\vint{\p}$ represents instead the interaction component of the velocity, which accounts for intercellular adhesive/repulsive interactions defined by kernels $H^{\p\q}:\Rn\to\Rn$ and performed by the test cell $X^\p_t$ through a scanning of the surrounding mass $\mu^\q_t$ (\ie, of its \emph{field cells}). Concerning this, the right way to interpret the specific subscripts is the following: the generic interaction kernel $H^{\p\q}$ defines how the test individual of population $\p$ \emph{is influenced} in its dynamics \emph{by} a field individual belonging to population $\q$. In particular, we distinguish between \emph{endogenous} (\ie, among individuals of the same population, $\p=\q$) and \emph{exogenous} (\ie, among individuals of different populations, $\p\neq\q$) contributions. It is finally natural to assume that the result of the intercellular interactions gives an effect along the direction connecting the interacting individuals, depending for instance on their relative position.

\subsection{Formal derivation via transport of measures}
\label{sec:formal_derivation}
The phenomenological model~\eqref{eq:test_cell} involves two types of state variables, $X^\p_t$ and the $\mu^\q_t$'s, so far correlated at a conceptual but not yet formal level. A formal link can be established via the following argument: the cell mass $\mu^\p_t(E)$, as a material quantity of a fixed number of cells of the $\p$-th population, has to move coherently with the movement of its single components. In mathematical terms, this implies that each $\mu^\p_t$ is transported, in a conservative manner, by the corresponding $X^\p_t$. Given an initial mass distribution $\mu^\p_0$, at all successive times $t>0$ it results therefore:
\begin{equation}
	\mu^\p_t(E)=\mu^\p_0(E_0), \quad \forall\,E\in\mathcal{B}(\Rn),
	\label{eq:pushfwd_a}
\end{equation}
where
\begin{equation}
    E_0:={(X^\p_t)}^{-1}(E)=\{x\in\Rn\,:\,X^\p_t(x)\in E\}
    \label{eq:pushfwd_b}
\end{equation}
is the set of all possible initial positions from which a test cell has reached the set $E$ at time $t>0$. Notice that if $X^\p_t$ is a Borel mapping then $E_0\in\mathcal{B}(\Rn)$, thus both sides of~\eqref{eq:pushfwd_a} are well defined.

In transport theory the operation defined by~\eqref{eq:pushfwd_a}-\eqref{eq:pushfwd_b} is termed the \emph{push forward} of the measure $\mu^\p_0$ by the mapping $X^\p_t$ and is denoted by the symbol $\#$:
$$ \mu^\p_t=X^\p_t\#\mu^\p_0. $$
An equivalent manner of characterizing the action of the push forward operator $\#$ is:
\begin{equation}
	\int_{\Rn}\varphi(x)\,d\mu^\p_t(x)=\int_{\Rn}\varphi(X^\p_t(x))\,d\mu^\p_0(x)
	\label{eq:pushfwd}
\end{equation}
for every bounded and Borel function $\varphi:\Rn\to\R$.

Using~\eqref{eq:pushfwd} it is possible to formally derive from~\eqref{eq:test_cell} an evolution equation for each mass measure $\mu^\p_t$. First, let us technically regard the mapping $t\mapsto\mu^\p_t$ as a curve in the space of distributions, \ie, the dual space of the Banach space $\Cc$ containing the infinitely differentiable functions with compact support in $\Rn$. Then, after picking a test function $\varphi\in\Cc$, one can compute the time derivative of $\mu^\p_t$ proceeding like in Reynolds' Theorem:
\begin{align}
	\begin{aligned}[b]
		\frac{d}{dt}\duality{\mu^\p_t}{\varphi} &=
			\frac{d}{dt}\int_{\Rn}\varphi(x)\,d\mu^\p_t(x)=
				\frac{d}{dt}\int_{\Rn}\varphi(X^\p_t(x))\,d\mu^\p_0(x)= \\
		&= \int_{\Rn}\nabla\varphi(X^\p_t(x))\cdot\dot{X}^\p_t(x)\,d\mu^\p_0(x)= \\
		&= \int_{\Rn}\nabla\varphi(X^\p_t(x))\cdot\left(\vchem{\p}(t,\,X^\p_t(x))+\vint{\p}(t,\,X^\p_t(x))\right)\,d\mu^\p_0(x)= \\
		&= \int_{\Rn}\nabla\varphi(x)\cdot\left(\vchem{\p}(t,\,x)+\vint{\p}(t,\,x)\right)\,d\mu^\p_t(x),
    \label{reynolds}
\end{aligned}
\end{align}
where $\duality{\cdot}{\cdot}$ denotes the duality pairing in $\Cc$. After defining the \emph{transport velocity}
\begin{equation}
	v^\p[\{\mu^\q_t\}_{\q=1}^{\PP}](t,\,x):=\vchem{\p}(t,\,x)+\vint{\p}(t,\,x)=
		\vchem{\p}(t,\,x)+\sum_{\q=1}^{\PP}\int_{\Rn\setminus\{x\}}H^{\p\q}(y-x)\,d\mu^\q_t(y),
	\label{eq:v}
\end{equation}
Eq.~\eqref{reynolds} can be recognized as a weak form of the conservation law for the cellular mass, leading to the corresponding strong formulation:
$$ \partial_t\mu^\p_t+\div(\mu^\p_t v^\p[\{\mu^\q_t\}_{\q=1}^{\PP}])=0, \quad \p=1,\,\dots,\,\PP. $$

The above-presented approach is expected to provide consistent mathematical tools to address the multiscale issues posed by biological problems. In fact, the description of the system in terms of an abstract measure does not force an \emph{a priori} choice of the scale of representation of each population, which can then be deferred to the specific structure given to $\mu^\p_t$. For instance, a \emph{macroscopic/continuous} description of a cellular population requires to assume that the mass measure is absolutely continuous with respect to the $n$-dimensional Lebesgue measure $\Leb{n}$ (\ie, $\mu^\p_t\ll\Leb{n}$):
$$ \mu^\p_t\equiv\rho^\p_t\Leb{n} \qquad \text{or} \qquad
    		\mu^\p_t(E)=\int_{E}\rho^\p_t(x)\,dx, \quad \forall\,E\in\mathcal{B}(\Rn), $$
where $\rho^\p_t:\Rn\to[0,\,+\infty)$ is the \emph{density} of the $\p$-th cell population at time $t$, whose existence is asserted by the Radon-Nikodym's Theorem. Conversely, a \emph{microscopic/discrete} description, in which each cell is individually represented, can be achieved by using an atomic mass measure constituted by a sum of Dirac masses centered in each cell position:
$$ \mu^\p_t\equiv\epsilon^\p_t:=\sum_{k=1}^{N^\p}\delta_{x^\p_k(t)} \qquad \text{or} \qquad
	\mu^\p_t(E)\equiv\epsilon^\p_t(E)=\sum_{k=1}^{N^\p}\delta_{x^\p_k(t)}(E), \quad \forall\,E\in\mathcal{B}(\Rn), $$
where the $x^\p_k$'s are the positions where cells are located at time $t$, \ie, $x^\p_k(t)=X^\p_t(x^\p_k(0))$.

The proposed modeling framework differentiates our point of view from other approaches widely studied in the literature, which generally interpret the relationship between discrete and continuous representations in an average sense by means of statistical distributions of mass-vanishing particles (the number of particles is sent to infinity while the total mass of the system is kept finite). Rather, we understand the discrete and continuous scales of description as, in principle, \emph{phenomenologically} different, since, in both cases, the total cellular mass is maintained finite. In other words, we are not describing the continuum as emerging in the limit $N^\p\to\infty$ from a vanishing discrete representation. Continuous and discrete descriptions are two complementary ways of looking at the same finite-size cell population, which can be associated to complementary phenomenologies exhibited by cells. The model also allows us to use, within the \emph{same} framework, a different type of representation for each population forming the biological system of interest.

\section{A two-population system}
\label{sec:two_pop}
In a wide range of pattern formations, characteristic of biological processes, large aggregates of non-specialized inactivated cells are collectively guided by a small number of specialized and activated individuals (often belonging to the same cell lineage). This happens for instance in angiogenesis, morphogenesis, and wound healing mechanisms, or in the metastatic infiltration of solid tumors~\cite{Friedl2009,Ilina2009}. On the one hand, such differentiated cells need to be properly described by a concentrated (\ie, discrete) distribution. On the other hand, the undifferentiated cell aggregates can be represented, on the whole, as a continuously distributed group with a low individual level of detail.

In order to formalize the mathematical description of such a case, we consider a two-population system ($\PP=2$) where the first population, whose mass distribution at time $t$ is denoted by $\mu^1_t$, is composed by a possibly high number $N^1$ of unspecialized cells whereas the second one, whose mass distribution at time $t$ is denoted by $\mu^2_t$, comprises a much lower number $N^2$ of differentiated cells. Applying the mathematical structures presented in Section~\ref{sec:math_model}, we are led to consider the following system of equations:
\begin{equation}
	\begin{cases}
		\partial_t\mu^1_t+\div(\mu^1_t v^1[\mu^1_t,\,\mu^2_t])=0 \\[2mm]
		\partial_t\mu^2_t+\div(\mu^2_t v^2[\mu^1_t,\,\mu^2_t])=0,
	\end{cases}
	\label{eq:2pop}
\end{equation}
which, regarding $\mu^1_t,\,\mu^2_t$ as distributions over the space of test functions $\Cc$, is properly understood in weak sense as:
$$ \frac{d}{dt}\duality{\mu^\p_t}{\varphi}+\duality{\div(\mu^\p_t v^\p[\mu^1_t,\,\mu^2_t])}{\varphi}=0,
	\quad \p=1,\,2, \quad \forall\,\varphi\in\Cc. $$
Using the divergence theorem, integrating in time, and taking into account that $\duality{\mu^\p_t}{\varphi}=\int_{\Rn}\varphi\,d\mu^\p_t$ (owing to Riesz's Representation Theorem) we finally get
\begin{equation}
	\int_{\Rn}\varphi\,d\mu^\p_t=\int_{\Rn}\varphi\,d\mu^\p_0
		+\int_0^t\int_{\Rn}\nabla\varphi\cdot v^\p[\mu^1_s,\,\mu^2_s]\,d\mu^\p_s\,ds,
			\quad \p=1,\,2, \quad \forall\,\varphi\in\Cc,
	\label{eq:2pop_weak}
\end{equation}
$\mu^\p_0$ being a prescribed initial mass distribution of either population.

In particular, by representing the unspecialized cells of population $1$ by a density $\rho^1_t$:
$$ \mu^1_t\equiv\rho^1_t\Leb{n}, \qquad \int_{\Rn}\rho^1_t(x)\,dx=N^1 $$
and the differentiated cells of population $2$ by atoms $\{x^2_k(t)\}_{k=1}^{N^2}$:
$$ \mu^2_t\equiv\epsilon^2_t:=\sum_{k=1}^{N^2}\delta_{x^2_k(t)} $$
and then plugging such measures in~\eqref{eq:2pop_weak} we obtain the following system of \emph{coupled} discrete-continuous equations, which consists of a conservation equation for $\rho^1_t$ and of a set of ODEs for the $x^2_k$'s:
\begin{equation}
	\begin{cases}
		\partial_t\rho^1_t+\div(\rho^1_t v^1[\rho^1_t,\,\epsilon^2_t])=0 \\[2mm]
		\dot{x}^2_k=v^2[\rho^1_t,\,\epsilon^2_t](t,\,x^2_k), \quad k=1,\,\dots,\,N^2,
	\end{cases}
	\label{eq:transport.syst}
\end{equation}
the velocity fields resulting from~\eqref{eq:v} as:
\begin{align}
	\left\{
	\begin{aligned}[c]
		& v^1[\rho^1_t,\,\epsilon^2_t](t,\,x)=\int_{\Rn}H^{11}(y-x)\rho^1_t(y)\,dy+\sum_{k=1}^{N^2}H^{12}(x^2_k-x) \\
		& v^2[\rho^1_t,\,\epsilon^2_t](t,\,x^2_k)=\vchem{2}(t,\,x^2_k)+\int_{\Rn\setminus\{x^2_k\}}H^{21}(y-x^2_k)\rho^1_t(y)\,dy
				+ \sum_{\substack{h=1 \\ h\neq k}}^{N^2}H^{22}(x^2_h-x^2_k).
	\end{aligned}
	\right.
	\label{eq:velocity_fields}
\end{align}

It is assumed that the chemotactic component of the velocity is present only for differentiated cells, which are sensitive to chemicals since they typically express surface receptors. In particular, $\vchem{2}$ depends on the local gradient of a molecular variable $c$:
\begin{equation}
	\vchem{2}(t,\,x)=k_0\nabla{c}(t,\,x),
	\label{eq:vchem2}
\end{equation}
where $k_0>0$ is a chemotactic strength. The chemical substance is assumed to evolve according to a standard reaction-diffusion equation:
\begin{equation}
	\partial_t c=D\Delta{c}+\alpha-\frac{c}{\tau}, \quad x\in\Rn,\ t\in(0,\,T]
	\label{eq:chemical}
\end{equation}
where $D$ is the constant and homogeneous diffusion coefficient and $1/\tau$ the decay rate. The secretion rate $\alpha=\alpha(t,\,x)$ describes where the chemical substance is produced.

\begin{figure}[!t]
\centering
\includegraphics[width=0.7\textwidth]{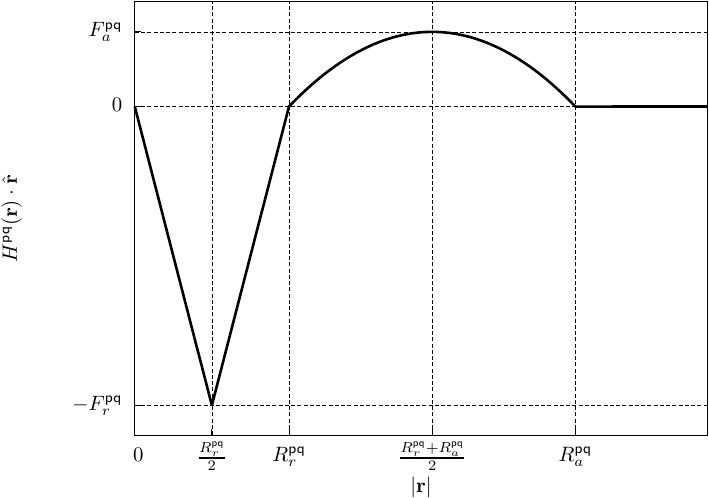}
\caption{Plot of the proposed interaction kernels $H^{\p\q}$, cf.~\eqref{eq:kernel}. $R^{\p\q}_r,\,R^{\p\q}_a$ are the repulsive and attractive radii, respectively, and $F^{\p\q}_r,\,F^{\p\q}_a$ the maximum strengths of the corresponding interactions.}
\label{fig.kernels}
\end{figure}

Finally, assuming isotropic adhesive/repulsive cell interactions, the kernels $H^{\p\q}$ are set to have the following form (see Fig.~\ref{fig.kernels}):
\begin{equation}
	H^{\p\q}(\rr)=
	\begin{cases}
		-2\dfrac{F^{\p\q}_r}{R^{\p\q}_r}\rr & \text{if\ } \abs{\rr}<\dfrac{R^{\p\q}_r}{2} \\[5mm]
		\left(2\dfrac{F^{\p\q}_r}{R^{\p\q}_r}\abs{\rr}-2F^{\p\q}_r\right)\hat{\rr} & \text{if\ } \dfrac{R^{\p\q}_r}{2}\leq\abs{\rr}<R^{\p\q}_r \\[5mm]
		-4F^{\p\q}_a\dfrac{(\abs{\rr}-R^{\p\q}_r)(\abs{\rr}-R^{\p\q}_a)}{(R^{\p\q}_r-R^{\p\q}_a)^2}\hat{\rr} &
			\text{if\ } R^{\p\q}_r\leq\abs{\rr}<R^{\p\q}_a \\[5mm]
		0 & \text{otherwise}
	\end{cases}
	\qquad\qquad \hat{\rr}:=\frac{\rr}{\abs{\rr}}
	\label{eq:kernel}
\end{equation}
where $R^{\p\q}_r,\,R^{\p\q}_a>0$ are the repulsive and adhesive radii and $F^{\p\q}_r, F^{\p\q}_a>0$ the corresponding interaction strengths, respectively. In the formula above $\rr,\,\hat{\rr}$ are the radial and unit radial vectors: the interactions expressed by the kernels $H^{\p\q}$ depend on the distance $\abs{\rr}$ between the interacting cells and are directed along the line $\hat{\rr}$ ideally connecting them.

Notice that, by fixing the interaction radii, we are introducing \emph{metric} intercellular interactions (\ie, within given maximum distances). Furthermore, in the case of two clones of the same cell line (\ie, population $2$ is the activated counterpart of population $1$), we can consistently take $R^{\p\q}_a=R_a$, $R^{\p\q}_r=R_r$ and $F^{\p\q}_r=F_r$ for all $\p,\,\q=1,\,2$.

\section{Numerical simulations}
\label{sec:simulations}
In this section we perform numerical simulations produced by a modified version of the numerical scheme developed in~\cite{Cristiani2011} (see also~\cite{Piccoli2013,Piccoli2011,Tosin2011} for error and convergence analysis), specifically designed for integrating the evolution of a diffusing substance within the basic measure-theoretic framework.

We deal with a two-dimensional ($n=2$) square bounded domain $\Omega\subset\R^2$, which is set to represent a $900\times 900\unit{{\mu m}^2}$ section of a Petri dish, typically used for \emph{in vitro} biological/biomedical assays. The computational time step $\Delta t$ is assumed to correspond to approximately $10$ seconds. Simulations are typically stopped after $T=3600\Delta{t}$, so that they reproduce a time-lapse of nearly $12$ hours.

In all realizations, the initial configuration of the system consists of few differentiated cells belonging to population $2$ distributed on the edge of a colony of inactivated individuals belonging to population $1$. The number $N^2$ of activated individuals will change in the different sets of simulations, whereas the number of cells within the continuous aggregate is fixed to $N^1=100$. In particular, population $1$ is initially arranged into a homogeneous and round colony, so that
$$
\rho^1_0(x)=
	\begin{cases}
		\dfrac{100}{\pi R_0^{2}} & \text{if\ } x\in B_{R_0} \\[3mm]
		0 & \text{otherwise}
	\end{cases}
$$
where $B_{R_0}\subset\Omega$ is a ball centered in the middle of the domain and $R_0=100\unit{\mu m}$ is its initial radius. This type of configuration may reproduce a tumor spheroid with an external ring of highly metabolic metastatic cells and an inner core of quiescent individuals, or an endothelial/epithelial cell system where few tip cells are able to act as guidance leaders for migration for the rest of the stalk/follower mass.

The parameters introduced in the interaction kernels~\eqref{eq:kernel}, describing the biophysical properties of the cells, can be evaluated consistently with biological considerations. In particular, the common repulsive radius $R_r$ is fixed equal to $20\unit{\mu m}$ (that is the average size of most eukaryotic cells) while the attractive one $R_a$ (that represents the maximal extension of cell filopods) can reach $60\unit{\mu m}$, i.e., three times the repulsive radius $R_r$. The intercellular repulsive strength $F_r$ is taken equal to $1\unit{\mu m\cdot s^{-1}}$. Conversely, the magnitude of the adhesive forces will change to reproduce different quantities of expressed cell adhesion molecules (CAMs). The microenvironmental chemical is assumed to diffuse and decay within the entire virtual Petri dish. In particular, referring to~\eqref{eq:chemical}, we set $\alpha=0$, $D=10\unit{\mu m^{2}\cdot s^{-1}}$ and $\tau=5000\unit{s}$: these are typical values for some important growth factors, such as many isoforms of the Vascular Endothelial Growth Factor (VEGF). In addition, we consider a production of chemical along a portion of the boundary of the domain, say $\Gamma\subseteq\partial\Omega$, where we prescribe:
$$ c=0.25\unit{\mu M\cdot s^{-1}} \quad \text{on\ } \Gamma. $$
In the remaining part we prescribe instead a no-flux condition:
$$ \nabla{c}\cdot\mathbf{n}=0 \quad \text{on\ } \partial\Omega\setminus\Gamma, $$
$\mathbf{n}$ being the outward unit normal to the boundary. Finally, the strength of the chemotactic term of population $2$, \ie, $k_{0}$, will be established in the following.

For a detailed analysis of the role played by the different model parameters, as well as for the corresponding sensitivity analysis, we refer to~\cite{Colombi}.

\subsection{Symmetric chemical field}
Let us investigate the evolution of the aforesaid cell system in the case of a symmetric chemical field, obtained by assuming that a constant production of the substance occurs on the entire boundary of the domain, \ie, $\Gamma\equiv\partial\Omega$.

In particular, in this first set of simulations we analyze cellular dynamics for different numbers of discrete individuals $N^2$. In these cases, the chemotactic strength is $k_0=500\unit{\mu m^2\cdot\mu M^{-1}\cdot s^{-1}}$, while the matrix collecting the adhesive interaction strengths is set to
\begin{equation}
    F_{a}=
    \begin{pmatrix}
    		0.04 & 0.203 \\
        0.0038 & 0.00025
    \end{pmatrix}
    \unit{\mu m\cdot s^{-1}}.
    	\label{eq:adh_force}
\end{equation}
The value of $F^{11}_a$ assures that population 1 reaches a sort of inner equilibrium, \ie, the continuous colony does not undergo a dramatic scatter or an unrealistic collapse (see~\cite{Colombi} for a detailed analysis). Moreover, this set of parameters allows for a consistent chemotactic migration of the activated cells, as the heterotypic adhesive forces $F^{12}_a$ and $F^{21}_a$ are not strong enough to cause their absorption within the continuous mass.

\begin{figure}[t!]
\centering
\includegraphics[width=\textwidth]{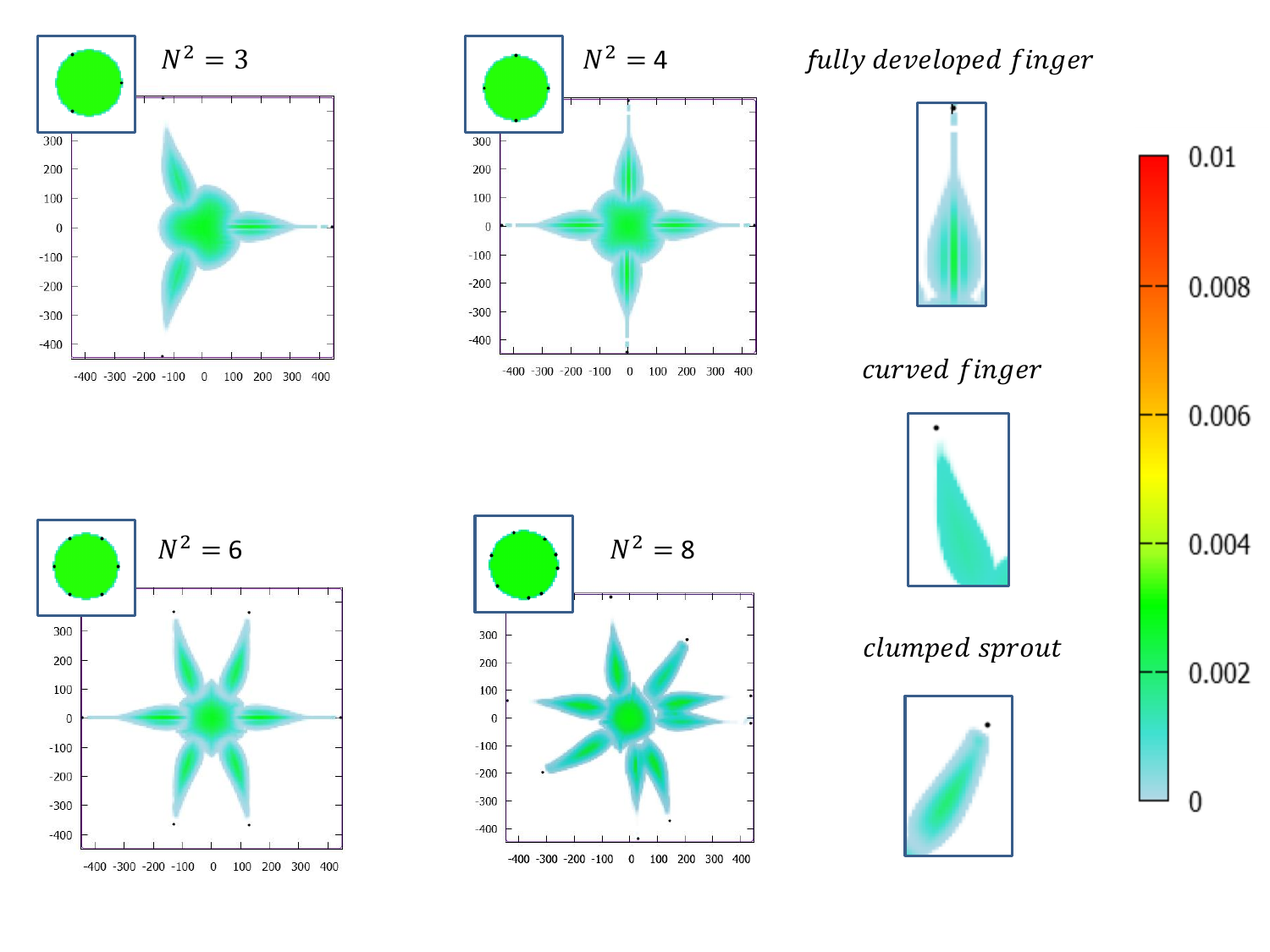}
\caption{Final configurations (\ie, at $T=3600\Delta{t}\approx 12\unit{h}$) of the two-population system for different numbers of activated cells. As represented in the insets, all initial configurations consist of a round aggregate of undifferentiated individuals surrounded by a ring of discrete cells. The parameter setting is described along the text, in particular, the adhesive strengths are summarized in~\eqref{eq:adh_force}.}
\label{fig:different_N2}
\end{figure}

\begin{figure}[t!]
\centering
\includegraphics[width=1\textwidth]{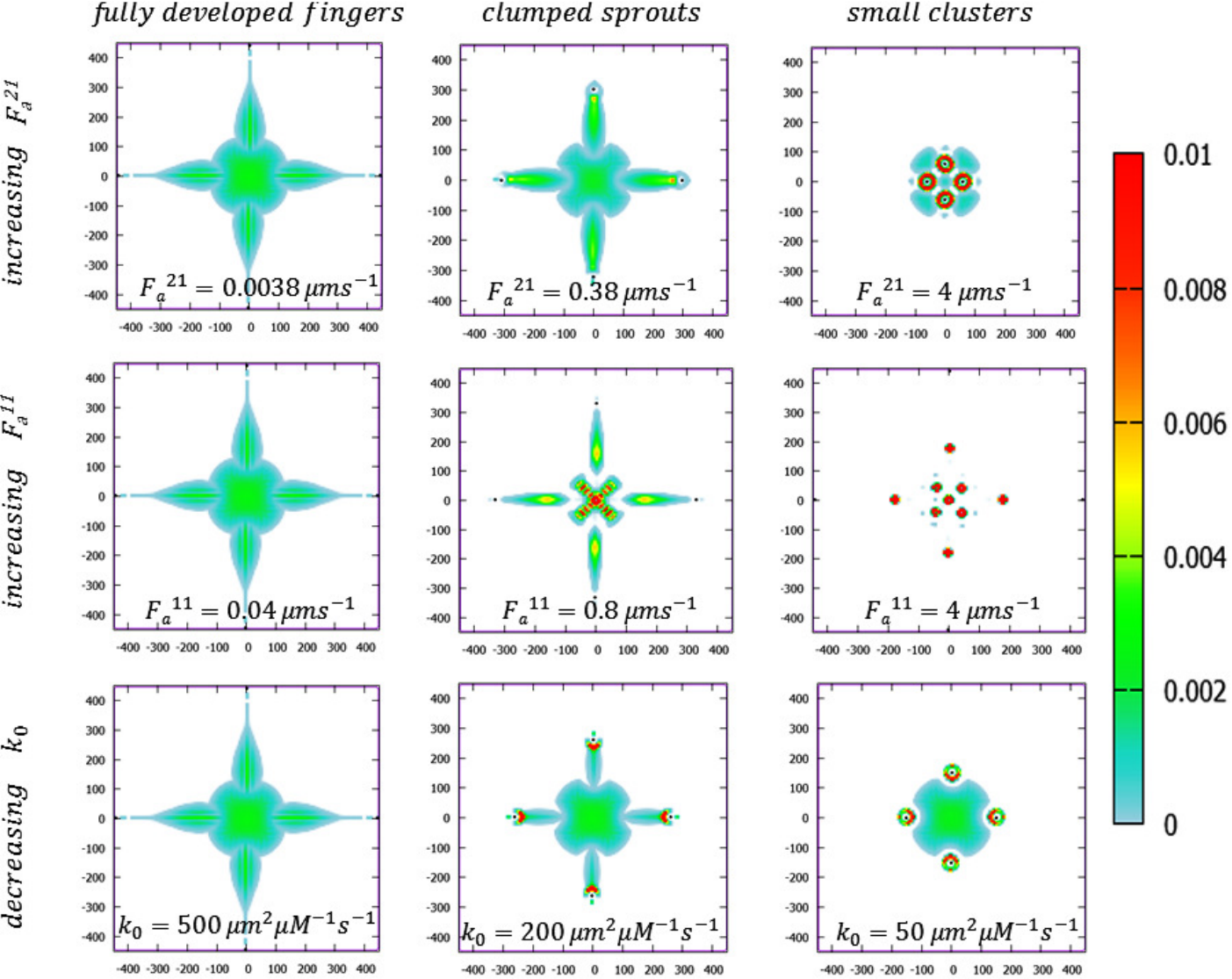}
\caption{Final configurations (\ie, at $T=3600\Delta{t}\approx 12\unit{h}$) of the two-population system for different values of either the adhesive strengths $F^{21}_a,\,F^{11}_a$, or the chemotactic sensitivity $k_0$. All other parameters, which do not vary with respect to the simulations of Fig.~\ref{fig:different_N2}, are clarified along the text.}
\label{fig:different_F_k}
\end{figure}

Further, we can preliminarily observe that, given the specific assumptions on its production, the concentration of the chemical has a smooth square-shaped profile in the proximity of the boundary of the domain, while stabilizing in a more round configuration in the middle. For this reason, the chemotactic gradient is higher along the axes than along the diagonal of the square domain. By analyzing Fig.~\ref{fig:different_N2}, it is then possible to observe the formation of different, often coexisting, types of morphological structures, that extend from the continuous colony beside each differentiated individual, being determined by their initial positions:
\begin{itemize}
\item cellular \emph{fingers} emerge and fully develop from the undifferentiated aggregate in correspondence of discrete cells initially placed in the direction of the center of the nearest edge;
\item \emph{clumped sprout} grow instead from population 1 behind discrete cells initially placed in correspondence of a diagonal of the domain. This phenomenon results from the fact that such activated individuals sense a chemotactic field substantially lower, and therefore they slowly move from the continuous aggregate, so that continuous fingers have not enough time to completely develop before the end of the observation time;
\item \emph{curved} but fully developed \emph{fingers} emerge from the undifferentiated colony $1$ in the case of discrete cells placed in intermediate positions between the direction of a corner of the domain and the direction of the central part of a domain edge. This is due to the fact that, in this case, the activated individuals quickly turn during migration, in order to align with the direction of maximal chemical gradient.
\end{itemize}
The growth rate of the different morphological phenotypes, \ie., nearly $30\unit{\mu m\cdot h^{-1}}$ in the case of fully developed (even curved) fingers and nearly $15\unit{\mu m\cdot h^{-1}}$ in the case of clumped sprouts, is determined by the speed of the corresponding leader cell. It is interesting to observe that these values are within the range of the typical speed measured for most endothelial cell lines (also of malignant origin).

Fixing $N^2=4$, we now turn to investigating the role played by some relevant model parameters. An increment of two orders of magnitude in the value of $F^{21}_{a}$ (\ie, $=0.38\unit{\mu m\cdot s^{-1}}$), which as seen measures the attractive potential exerted \emph{by} the undifferentiated cells \emph{on} the discrete individuals, results in a partial disruption of the fingering process, see Fig.~\ref{fig:different_F_k} (top-center panel). In fact, only clumped sprouts emerge, which are characterized by a higher cellular density but a lower growth rate with respect to fully developed fingers. This can be explained observing that the chemotactic migration of the activated cells is partially inhibited by the increased intercellular adhesive interactions. Finally, for $F^{21}_{a}>1\unit{\mu m\cdot s^{-1}}$ the activated individuals are absorbed within the continuous colony, which undergoes an inner reorganization, possibly stabilizing into a four cluster-morphology (with each cluster containing a discrete cell), as reproduced in the top-right panel of Fig.~\ref{fig:different_F_k}. This phenomenon is due to the fact that the exogenous attractive force  is high enough to prevent at all the chemotactic movement of the leader individuals, which are therefore attracted more by the surrounding undifferentiated mass than by the chemical source.

A similar influence on the evolution of the cellular system can be observed in the case of increments in the endogenous adhesive strength of the population $1$, see Fig.~\ref{fig:different_F_k} (middle panels). In fact, at $F^{11}_a\approx 1\unit{\mu m\cdot s^{-1}}$ four thin sprouts detach from the continuous mass and follow the discrete leaders. In particular, such motile structures are characterized by a higher cellular concentration in the trailing edge. The remaining part of population $1$ instead undergoes a significant morphological transition toward a cross-like shape. At significantly larger values of $F^{11}_a$ (\ie, $> 1\unit{\mu m\cdot s^{-1}}$) the differentiated cells possibly escape the continuous spheroid, which remains still and reorganizes into nine small and dense clusters. It is interesting to notice that these little cellular islands are exactly located at the center of the initial colony and, referring to the previous case, at the extremes of the cross-like shape and in the center of the dense trailing part of the emerging sprouts.

We finally vary the chemotactic strength of the discrete cells, see Fig.~\ref{fig:different_F_k} (bottom panels). A drop of $k_0$ from $500\unit{\mu m^2\cdot \mu M^{-1}\cdot s^{-1}}$ to $200\unit{\mu m^2\cdot \mu M^{-1}\cdot s^{-1}}$ results in a partial inhibition of the fingering process. A decrement of the relative importance of the directional velocity for the activated individuals, accompanied by the relative increment of the importance of the exogenous adhesive interactions, leads to the (slow) formation of thin and short sprouts, that are characterized by a higher cellular density in their frontal part, \ie, around the leader cells. At even lower values of $k_0$ (\ie, $\leq 100\unit{\mu m^2\cdot\mu M^{-1}\cdot s^{-1}}$), all discrete individuals are retained within small clusters stabilized in the proximity of the boundary of the non-motile continuous mass, whose bulk remains, in turn, in a non-motile quasi-rigid configuration. The chemotactic force is in fact no longer able to overcome the exogenous adhesive interactions.

In summary, we can distinguish three different phenotypes of cell dynamics, each of them characteristic of a specific parameter range: the fingering process, which emerges in the case of low values of both $F^{21}_a$ and $F^{11}_a$ and/or of high values of $k_0$; the formation of short and clumped sprouts, obtained by increasing such adhesive strengths and/or by decreasing the chemotactic strength; and, finally, the aggregate clusterization into small islands, characteristic of dramatically high values of $F^{21}_a$ and $F^{11}_a$ and/or of low enough values of $k_0$.

\subsection{Asymmetric chemical field}

\begin{figure}[t!]
\centering
\includegraphics[width=\textwidth]{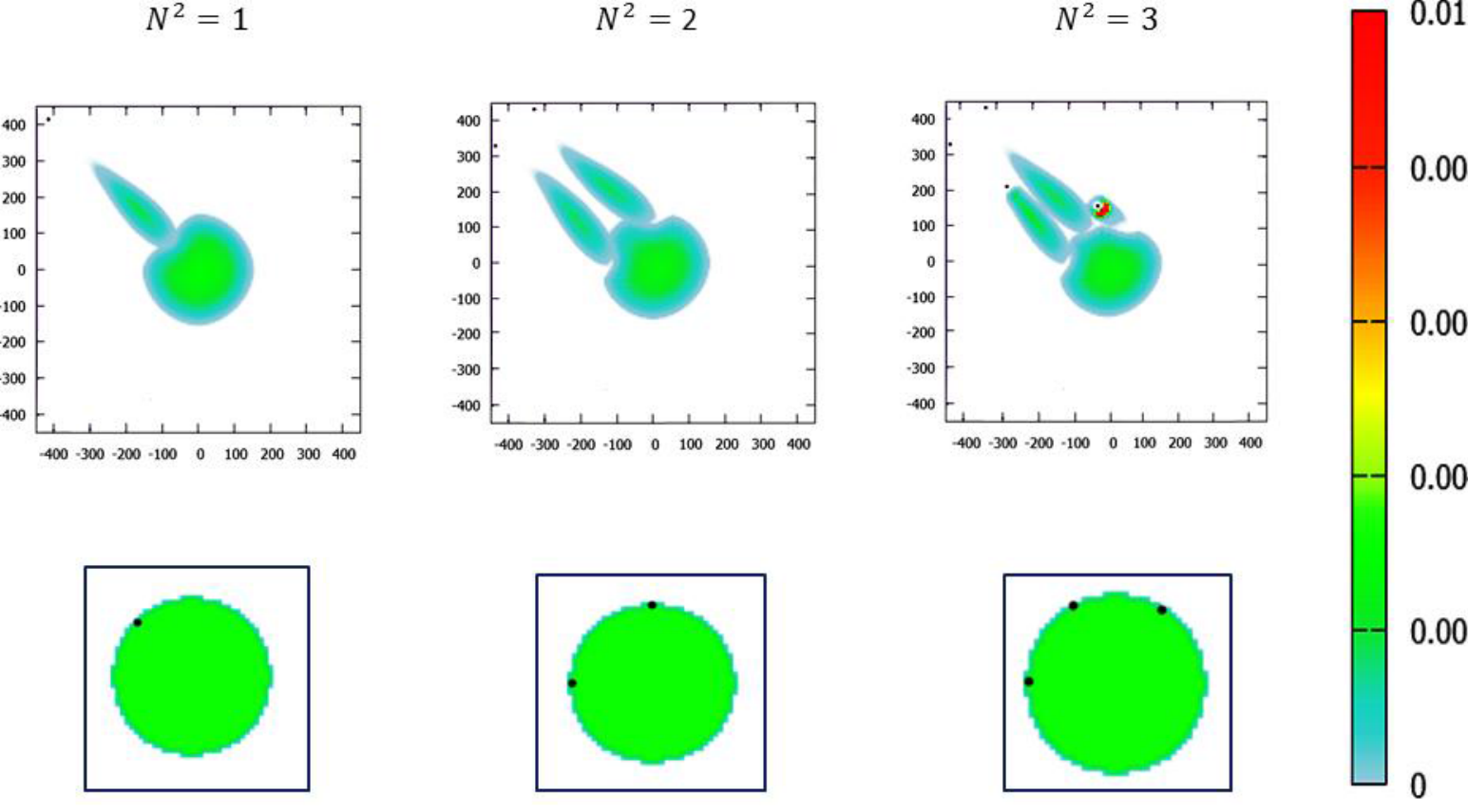}
\caption{Final configurations (\ie, at $T=3600\Delta{t}\approx 12\unit{h}$) of the two-population system for different numbers and locations of the activated cells, in the case of an asymmetric chemical field obtained by assuming that the chemical production occurs only at the top-left part of the boundary of the domain. The bottom insets represent the initial configurations of the cellular aggregate. The parameter setting is the same as that of the simulations presented in Fig.~\ref{fig:different_N2}.}
\label{fig:different_N2_asymm_c}
\end{figure}

We now turn to study the case of an angular source of chemical field, as the diffusive substance is only produced near the upper-left corner of the domain, \ie,
$$ \Gamma=\{(x,y)\in\partial\Omega:x=-450\unit{\mu m},\,y\in[400,450]\unit{\mu m}\ \textup{or\ }
	x\in[-450,-400]\unit{\mu m},\,y=450\unit{\mu m}\}. $$
All the parameters regulating the evolution of the chemical factor are the same as in the previous sets of simulations. The initial configuration of the continuous aggregate is again a homogeneous round colony of $N^1=100$ individuals, whereas we vary the number and the location of the discrete cells distributed around the undifferentiated mass. The adhesive intercellular strengths are defined like in~\eqref{eq:adh_force}, whereas the chemotactic sensitivity is taken as $k_0=500\unit{\mu m^2\cdot\mu M^{-1}\cdot s^{-1}}$.

As reproduced in Fig.~\ref{fig:different_N2_asymm_c} (left panels), in the case of a single activated individual, placed in the direction of the angle of the domain, a finger forms and extends from the continuous aggregate. Such a cellular structure grows as quickly as the movement of the discrete individual toward the source of the chemical. Almost the same phenomenon happens when two discrete cells are symmetrically distributed with respect to the direction of the chemical source, although the two emerging fingers are a bit bended (see the central panels of Fig.~\ref{fig:different_N2_asymm_c}).

It is even more interesting the case of three discrete cells symmetrically distributed at the top-left edge of the continuous colony, see Fig.~\ref{fig:different_N2_asymm_c} (right panels). In fact, we can observe the simultaneous formation of the three morphological phenotypes encountered in the previous section: a fully developed finger grows behind the ``central'' activated individual, a shorter sprout follows the discrete cell ``on the left'', and a partial clusterization of the undifferentiated aggregate is localized around the discrete cell ``on the right''. These dynamics are not surprising, as each discrete cell senses a distinct chemotactic gradient (we recall that the diffusive substance is produced only at the top-left corner of the boundary of the domain). In more detail, we can compare such different phenomenologies to those obtained with variations of the chemotactic coefficient $k_0$ in the case of the symmetric chemical profile (cf. Fig.~\ref{fig:different_F_k}, bottom panels). In particular, the ``central'' activated individual is subject to the maximum gradient, which corresponds for instance to the maximum value of $k_0$ in the previous set of simulations. On the opposite, the differentiated cell ``on the right'' senses the minimum gradient: this is consistent with the case of the minimum chemotactic strength. Finally, the activated cell ``on the left'' of the colony senses a gradient of an intermediate steepness, therefore only a clumped sprout can emerge, which is exactly the same cellular structure observed in the case of the intermediate value of $k_0$.

\section{Comparison of discrete and continuous descriptions}
\label{sec:comparison}
In this section we further investigate the discrete and continuous representations introduced in the setup of two interacting cell populations. While it is clear that population $2$, composed by few cells, has to be represented as a pointwise (viz. discrete) mass distribution, the question arises whether population $1$ needs a continuous or a discrete representation. Both can in principle be appropriate, even if the phenomenology of the problem naturally leads to choose the continuous description. The purpose of this section is to inquire, using analytical methods, into the similarities and differences in the dynamics of the coupled system caused by switching from the continuous to a conceivable discrete representation of the first population. In particular, we will use $\Rn$ as the reference domain, since we aim at investigating the qualitative role of some phenomenological parameters of the model, such as the strength of cell-cell interactions and the interaction radii, aside from technicalities which would derive from the presence of boundaries. Finally, it is useful to recall that the evolution equation~\eqref{eq:chemical} of the chemical field is uncoupled from the system of equations~\eqref{eq:2pop} regulating cell dynamics.

We mention that in~\cite{DiFrancesco2013} the authors deal with existence and uniqueness of solutions to systems of evolving mass measures which share some analogies with those presented in this paper. In particular, they consider two-population systems featuring self- and cross-interactions in the absence of guidance contributions that may arise from the dynamics of external fields. Interested readers are referred to that paper for such aspects of the theory. Instead, as anticipated before, the focus here is on obtaining estimates for comparing continuous and discrete descriptions of the same physical mass while using different spatial representations for coexisting but functionally distinct particle subsystems.

\subsection{Preliminaries}
\label{sec:preliminaries}
In this section we introduce the formulation of the problem and quickly review the tools which we need to study it.

We denote by $\cM_1^N(\Rn)$ the space of positive measures on $\Rn$ having finite mass $N$ and finite first moment:
$$ \cM_1^N(\Rn):=\left\{\mu:\mathcal{B}(\Rn)\to [0,\,+\infty)\ \text{s.t.\ } \mu(\Rn)=N,\ \int_{\Rn}\abs{x}\,d\mu(x)<+\infty\right\}. $$
We endow the space $\cM_1^N(\Rn)$ with the \emph{(first) Wasserstein metric}:
$$ W_1(\mu,\,\nu):=\sup_{\varphi\in\Lipone}\int_{\Rn}\varphi(x)\,d(\nu-\mu)(x), \quad \mu,\,\nu\in\cM_1^N(\Rn), $$
where $\Lipone$ denotes the space of Lipschitz continuous functions $\varphi:\Rn\to\R$ with at most unit Lipschitz constant. Usually $W_1$ is introduced in the context of probability measures, \ie, for measuring distances between measures with unit mass. However, its use for measures with arbitrary (but finite) mass according to the previous definition is straightforward.

Moreover, we define the space:
$$ \bM:=\cM_1^{N^1}(\Rn)\times\cM_1^{N^2}(\Rn) $$
of pairs of positive measures with masses $N^1,\,N^2$, respectively, and finite first moments. In $\bM$ we consider the metric:
$$ \bW_1((\mu^1,\,\mu^2),\,(\nu^1,\,\nu^2)):=W_1(\mu^1,\,\nu^1)+W_1(\mu^2,\,\nu^2) $$
for $\mu^1,\,\nu^1\in\cM^{N^1}(\Rn)$ and $\mu^2,\,\nu^2\in\cM^{N^2}(\Rn)$.

We finally introduce the space $\CTM$, $T>0$ being a final time, of continuous curves of measures in $\bM$ parameterized by time, $t\mapsto (\mu^1_t,\,\mu^2_t)\in\bM$, endowed with the metric
$$ \sup_{t\in[0,\,T]}\bW_1((\mu^1_t,\,\mu^2_t),\,(\nu^1_t,\,\nu^2_t)). $$

Following the notation introduced in Section~\ref{sec:two_pop}, we denote by $\rho^1_t$ and $\epsilon^1_t:=\sum_{k=1}^{N^1}\delta_{x^1_k(t)}$ the continuous and discrete descriptions of the mass of population $1$. The corresponding solution for population $2$ is then denoted by $\epsilon^{2,c}_t$, $\epsilon^{2,d}_t$, respectively.

\begin{remark}
We stress that the superscripts $c,\,d$ refer to the continuous or discrete nature of the \emph{first} population, the second one being always discrete by definition (as the symbol $\epsilon$ used for its mass measure is meant to recall). Notice that, for fixed $t$, we have $\rho^1_t,\,\epsilon^1_t\in\cM^{N^1}(\Rn)$ and $\epsilon^{2,c}_t,\,\epsilon^{2,d}_t\in\cM^{N^2}(\Rn)$.
\end{remark}

\begin{remark}
Starting from this section, for convenience of notation, we will systematically confuse the density $\rho^1_t$ with the measure that it represents. Therefore, given a function $\varphi:\Rn\to\R$, we will write e.g., $\int_{\Rn}\varphi(x)\,d\rho^1_t(x)$ to mean $\int_{\Rn}\varphi(x)\rho^1_t(x)\,dx$.
\end{remark}

Let $(\rho^1_0,\,\epsilon^2_0),\,(\epsilon^1_0,\,\epsilon^2_0)\in\bM$ be two possible initial configurations of the two-population system, in which population $1$ is either continuous or discrete while population $2$ is described in both cases by the same discrete distribution $\epsilon^2_0\in\cM^{N^2}(\Rn)$. Starting from such initial conditions, the Cauchy problem associated with system~\eqref{eq:2pop} (considered in the weak formulation~\eqref{eq:2pop_weak}) generates either solution $(\rho^1_t,\,\epsilon^{2,c}_t),\,(\epsilon^1_t,\,\epsilon^{2,d}_t)\in\CTM$. The goal is to study how much similar or dissimilar these solutions are depending on the similarity of the chosen initial mass distributions $\rho^1_0,\,\epsilon^1_0\in\cM^{N^1}(\Rn)$ of the first population. We anticipate that we will establish, under suitable assumptions specified below, an \emph{a priori} stability estimate of the form (cf. Theorem~\ref{theo:estimate})
\begin{equation}
	\sup_{[0,\,T]}\bW_1((\rho^1_t,\,\epsilon^{2,c}_t),\,(\epsilon^1_t,\,\epsilon^{2,d}_t))\leq\cC W_1(\rho^1_0,\,\epsilon^1_0),
	\label{eq:stability.est}
\end{equation}
$\cC>0$ being a constant depending on the parameters of the model. Such a result will allow us to compare the hybrid continuous-discrete solution $(\rho^1_t,\,\epsilon^{2,c}_t)$ with the conceivable fully discrete one $(\epsilon^1_t,\,\epsilon^{2,d}_t)$. More precisely, we will see that, in general, for fixed numbers $N^1,\,N^2$ of cells of the two populations, the similarity between the said solutions is established by some key parameters of the model.

\subsection{Regularity of the velocity field}
Estimate~\eqref{eq:stability.est} essentially relies on the regularity of the transport velocity fields $v^1,\,v^2$ given by~\eqref{eq:v},~\eqref{eq:velocity_fields}.

First, it is not difficult to see that the chemotactic part of $\vchem{2}$ is Lipschitz continuous in $x$, in the sense that there exists a constant $\Lip{\vchem{2}}>0$ such that:
$$ \abs{\vchem{2}(t,\,x_2)-\vchem{2}(t,\,x_1)}\leq\Lip{\vchem{2}}\abs{x_2-x_1},
	\quad \forall\,x_1,\,x_2\in\Rn,\ t\in (0,\,T]. $$
In fact, if we assume that the secretion rate $\alpha=\alpha(t,\,x)$ in~\eqref{eq:chemical} is continuous in time and piecewise continuous in space (including, in particular, the possibility that secretion occurs in specific regions of the domain and vanishes elsewhere), we obtain that the chemical field solving the reaction-diffusion equation is continuous in time and differentiable in space and that $\nabla{c}$ appearing in the definition of $\vchem{2}$, cf.~\eqref{eq:vchem2}, is continuous in time and in space (actually piecewise-differentiable in space). Notice that we do not need to take into account boundary conditions since we work in $\Rn$ as the reference domain.

Second, concerning the interaction part of the velocity, under the assumption of Lipschitz continuous kernels $H^{\p\q}$ (which we incidentally notice is satisfied by the kernels proposed in~\eqref{eq:kernel}), we have, for all $(\mu^1,\,\mu^2),\,(\nu^1,\,\nu^2)\in\bM$ and all fixed $x\in\Rn$,
\begin{align*}
	\vert\vint{\p}[\nu^1,\,\nu^2](x) &- \vint{\p}[\mu^1,\,\mu^2](x)\vert \\
	& \leq\abs{\int_{\Rn}H^{\p 1}(y-x)\,d(\nu^1-\mu^1)(y)}+\abs{\int_{\Rn}H^{\p 2}(y-x)\,d(\nu^2-\mu^2)(y)} \\
	& \leq\Lip{H^{\p 1}}W_1(\mu^1,\,\nu^1)+\Lip{H^{\p 2}}W_1(\mu^2,\,\nu^2) \\
	& \leq\max\{\Lip{H^{\p 1}},\,\Lip{H^{\p 2}}\}\bW_1((\mu^1,\,\mu^2),\,(\nu^1,\,\nu^2)).
\end{align*}
On the other hand, if $(\mu^1,\,\mu^2)\in\bM$ is fixed, for all $x_1,\,x_2\in\Rn$ it results:
\begin{align*}
	\abs{\vint{\p}[\mu^1,\,\mu^2](x_2)-\vint{\p}[\mu^1,\,\mu^2](x_1)} &\leq \int_{\Rn}\abs{H^{\p 1}(y-x_2)-H^{\p 1}(y-x_1)}\,d\mu^1(y) \\
	& \phantom{\leq} +\int_{\Rn}\abs{H^{\p 2}(y-x_2)-H^{\p 2}(y-x_1)}\,d\mu^2(y) \\
	& \leq (N^1\Lip{H^{\p 1}}+N^2\Lip{H^{\p 2}})\abs{x_2-x_1}.
\end{align*}
Collecting these results we can state finally:
\begin{proposition}[Regularity of the velocity fields]
Let the interaction kernels $H^{\p\q}$, $\p,\,\q=1,\,2$, be Lipschitz continuous in $\Rn$. Then the transport velocities $v^\p$, $\p=1,\,2$, given by~\eqref{eq:v},~\eqref{eq:velocity_fields} are Lipschitz continuous as well in $\Rn\times\bM$, i.e., there exist constants $\Lip{v^\p}>0$ such that:
$$ \abs{v^\p[\nu^1,\,\nu^2](t,\,x_2)-v^\p[\mu^1,\,\mu^2](t,\,x_1)}\leq
	\Lip{v^\p}\biggl(\abs{x_2-x_1}+\bW_1((\mu^1,\,\mu^2),\,(\nu^1,\,\nu^2))\biggr), \quad \p=1,\,2 $$
for all $x_1,\,x_2\in\Rn$, $(\mu^1,\,\mu^2),\,(\nu^1,\,\nu^2)\in\bM$, and fixed $t\in (0,\,T]$. Furthermore:
$$ \Lip{v^\p}\leq\max\left\{\Lip{\vchem{\p}}+\sum_{\q=1}^{2}N^\q\Lip{H^{\p\q}},\,\max_{\q=1,\,2}\Lip{H^{\p\q}}\right\}, \quad \p=1,\,2. $$
\label{prop:reg_v}
\end{proposition}

\begin{remark}
For $N^\p\geq 1$, $\p=1,\,2$, the estimate for $\Lip{v^\p}$ reduces to
$$ \Lip{v^\p}\leq\Lip{\vchem{\p}}+\sum_{\q=1}^{2}N^\q\Lip{H^{\p\q}}, \quad \p=1,\,2. $$
Moreover, since in the two-population setting under consideration we have $\vchem{1}\equiv 0$, in the estimate for $\Lip{v^1}$ it results actually $\Lip{\vchem{1}}=0$.
\end{remark}

\subsection{Notion of solution and trajectories}
As already defined in Section~\ref{sec:two_pop}, a pair $(\mu^1_\bullet,\mu^2_\bullet)\in\CTM$ is a weak solution to the non-autonomous system~\eqref{eq:2pop} if it satisfies~\eqref{eq:2pop_weak}. Moreover, one can formally check that
$$ (\mu^1_t,\,\mu^2_t)=(X^1_t\#\mu^1_0,\,X^2_t\#\mu^2_0), \quad t\in (0,\,T] $$
is a representation formula of the solution if the \emph{flow maps} $X^\p_t(x):\Rn\to\Rn$, $\p=1,\,2$ (cf. Section~\ref{sec:formal_derivation}), satisfy:
\begin{equation}
	\begin{cases}
		\partial_t X^\p_t(x)=v^\p[\mu^1_t,\,\mu^2_t](t,\,X^\p_t(x)) \\[1mm]
		X^\p_{0}(x)=x
	\end{cases}
	\label{eq:flow_maps}
\end{equation}
for $x\in\Omega$ and $t\in (0,\,T]$. Existence and uniqueness of such flow maps, and the fact that they provide representation formulas for the solutions to~\eqref{eq:2pop_weak}, follow from the theory developed in~\cite{Ambrosio2008}. In particular, $v^1$ and $v^2$ fulfill the required assumptions, being both Borel vector fields with global bounds.

Taking into account the Lipschitz continuity of $v^\p$, $\p=1,\,2$, asserted by Proposition~\ref{prop:reg_v} we now obtain the Lipschitz continuity of the flow maps~\eqref{eq:flow_maps}.
\begin{proposition}[Regularity of the flow maps] \hfill
\begin{enumerate}
\item[(i)] For $(\mu^1_\bullet,\,\mu^2_\bullet)\in\CTM$ the flow maps defined by~\eqref{eq:flow_maps} satisfy
$$ \abs{X^\p_t(x_2)-X^\p_t(x_1)}\leq e^{\Lip{v^\p}t}\abs{x_2-x_1}, \quad \p=1,\,2, \quad \forall\,x_1,\,x_2\in\Rn. $$
\item[(ii)] Let $x\in\Rn$ and $(\mu^1_\bullet,\,\mu^2_\bullet),\,(\nu^1_\bullet,\,\nu^2_\bullet)\in\CTM$. Call $X^{\p,\mu}_t,\,X^{\p,\nu}_t$, $\p=1,\,2$, the flow maps computed from~\eqref{eq:flow_maps} with either $(\mu^1_\bullet,\,\mu^2_\bullet)$ or $(\nu^1_\bullet,\,\nu^2_\bullet)$, respectively. Then:
$$ \abs{X^{\p,\nu}_t(x)-X^{\p,\mu}_t(x)}\leq \Lip{v^\p}e^{\Lip{v^\p}t}\int_0^t\bW_1((\mu^1_s,\,\mu^2_s),\,(\nu^1_s,\,\nu^2_s))\,ds, \quad \p=1,\,2. $$
\end{enumerate}
\label{prop:reg_gamma}
\end{proposition}
\begin{proof}
From~\eqref{eq:flow_maps} it follows $X^\p_t(x)=x+\int_0^t v^\p[\mu^1_s,\,\mu^2_s](s,\,X^\p_s(x))\,ds$. Using this formula we get:
\begin{enumerate}
\item[(i)]
$\begin{aligned}[t]
	\abs{X^\p_t(x_2)-X^\p_t(x_1)}	&\leq
		\abs{x_2-x_1}+\int_0^t\abs{v^\p[\mu^1_s,\,\mu^2_s](s,\,X^\p_s(x_2))-
			v^\p[\mu^1_s,\,\mu^2_s](s,\,X^\p_s(x_1))}\,ds \\
	&\leq\abs{x_2-x_1}+\Lip{v^\p}\int_0^t\abs{X^\p_s(x_2)-X^\p_s(x_1)}\,ds,
\end{aligned}$

where the second inequality follows from Proposition~\ref{prop:reg_v}. Invoking further Gronwall's inequality yields the thesis.
\item[(ii)]
$\abs{X^{\p,\nu}_t(x)-X^{\p,\mu}_t(x)}\leq\ds{\int_0^t}\abs{v^\p[\nu^1_s,\,\nu^2_s](s,\,X^{\p,\nu}_s(x))-
	v^\p[\mu^1_s,\,\mu^2_s](s,\,X^{\p,\mu}_s(x))}\,ds$

whence, owing to Proposition~\ref{prop:reg_v}, we discover:
$$ \abs{X^{\p,\nu}_t(x)-X^{\p,\mu}_t(x)}\leq
	\Lip{v^\p}\int_0^t\biggl(\abs{X^{\p,\nu}_s(x)-X^{\p,\mu}_s(x)}+\bW_1((\mu^1_s,\,\mu^2_s),\,(\nu^1_s,\,\nu^2_s))\biggr)\,ds, $$
and the proof is concluded by the application of Gronwall's inequality. \qedhere
\end{enumerate}
\end{proof}

\subsection{Main results}
Thanks to Proposition~\ref{prop:reg_gamma} we are now in a position to prove the main result of this part, which establishes that the discrepancy between the hybrid solution $(\rho^1_t,\,\epsilon^{2,c}_t)$ and its fully discrete counterpart $(\epsilon^1_t,\,\epsilon^{2,d}_t)$ introduced in Section~\ref{sec:preliminaries} is controlled \emph{a priori} by the discrepancy between the continuous and discrete versions of the initial condition of population $1$. Next we will also provide a way to estimate, at least for simple but representative configurations, the initial discrepancy between the two possible descriptions of population $1$.

The precise statement of the result is as follows:

\begin{theorem}[\emph{A priori} stability estimate]
Let $(\rho^1_\bullet,\,\epsilon^{2,c}_\bullet),\,(\epsilon^1_\bullet,\,\epsilon^{2,d}_\bullet)\in\CTM$ be the hybrid and fully discrete solutions to system~\eqref{eq:2pop}, understood in the weak formulation~\eqref{eq:2pop_weak}, corresponding to initial data\footnote{Recall that, by definition, population $2$ has the same initial condition in both cases.} $(\rho^1_0,\,\epsilon^2_0),\,(\epsilon^1_0,\,\epsilon^2_0)\in\bM$, respectively. Assume moreover that the interaction kernels $H^{\p\q}$, $\p,\,\q=1,\,2$, are Lipschitz continuous in $\Rn$. Then estimate~\eqref{eq:stability.est} holds true with a constant $\cC>0$ independent of $\rho^1_0,\,\epsilon^2_0$.
\label{theo:estimate}
\end{theorem}
\begin{proof}Let $\varphi\in\Lipone$, then:
\begin{align*}
	&\int_{\Rn}\varphi\,d(\epsilon^1_t-\rho^1_t)+\int_{\Rn}\varphi\,d(\epsilon^{2,d}_t-\epsilon^{2,c}_t) \\
	&=\int_{\Rn}\varphi\circ X^{1,d}_t\,d\epsilon^1_0-\int_{\Rn}\varphi\circ X^{1,c}_t\,d\rho^1_0
		+\int_{\Rn}(\varphi\circ X^{2,d}_t-\varphi\circ X^{2,c}_t)\,d\epsilon^2_0 \\
\intertext{where $X^{\p,c}_t,\,X^{\p,d}_t$, $\p=1,\,2$, denote the flow maps given by~\eqref{eq:flow_maps} when the first population is either continuous or discrete, respectively. Adding and subtracting $\int_{\Rn}(\varphi\circ X^{1,d}_t)\,d\rho^1_0$ we further find:}
	&=\int_{\Rn}\varphi\circ X^{1,d}_t\,d(\epsilon^1_0-\rho^1_0)+\int_{\Rn}(\varphi\circ X^{1,d}_t-\varphi\circ X^{1,c}_t)\,d\rho^1_0
		+\int_{\Rn}(\varphi\circ X^{2,d}_t-\varphi\circ X^{2,c}_t)\,d\epsilon^2_0.
\end{align*}

Next we observe that, owing to Proposition~\ref{prop:reg_gamma}(i), the mapping $(\varphi\circ X^{1,d}_t)(x):\Rn\to\R$ is Lipschitz continuous in $x$ with $\Lip{\varphi\circ X^{1,d}_t}\leq\Lip{X^{1,d}_t}\leq e^{\Lip{v^1}t}$. Therefore:
$$ \int_{\Rn}\varphi\circ X^{1,d}_t\,d(\epsilon^1_0-\rho^1_0)\leq e^{\Lip{v^1}t}W_1(\rho^1_0,\,\epsilon^2_0). $$

On the other hand, because of Proposition~\ref{prop:reg_gamma}(ii) we have:
\begin{align*}
	\abs{(\varphi\circ X^{\p,d}_t)(x)-(\varphi\circ X^{\p,c}_t)(x)} &\leq \abs{X^{\p,d}_t(x)-X^{\p,c}_t(x)} \\
	&\leq \Lip{v^\p}e^{\Lip{v^\p}t}\int_0^t\bW_1((\rho^1_s,\,\epsilon^{2,c}_s),\,(\epsilon^1_s,\,\epsilon^{2,d}_s))\,ds,
\end{align*}
whence, considering that $\rho^1_0(\Rn)=N^1$ and $\epsilon^2_0(\Rn)=N^2$,
\begin{align*}
	\int_{\Rn}(\varphi\circ X^{1,d}_t-\varphi\circ X^{1,c}_t)\,d\mu^\p_0 &\leq
		\int_{\Rn}\abs{\varphi\circ X^{1,d}_t-\varphi\circ X^{1,c}_t}\,d\mu^\p_0 \\
	&\leq N^\p\Lip{v^\p}e^{\Lip{v^\p}t}\int_0^t\bW_1((\rho^1_s,\,\epsilon^{2,c}_s),\,(\epsilon^1_s,\,\epsilon^{2,d}_s))\,ds,
		\quad \p=1,\,2
\end{align*}
with $\mu^1_0=\rho^1_0$, $\mu^2_0=\epsilon^2_0$.

Finally, collecting together the estimates obtained so far we discover:
\begin{align*}
	\int_{\Rn}\varphi\,d(\epsilon^1_t-\rho^1_t)+\int_{\Rn}\varphi\,d(\epsilon^{2,d}_t-\epsilon^{2,c}_t) &\leq
		e^{\Lip{v^1}t}W_1(\rho^1_0,\,\epsilon^2_0) \\
	&\phantom{\leq} +\left(\sum_{\p=1}^{2}N^\p\Lip{v^\p}e^{\Lip{v^\p}t}\right)
		\int_0^t\bW_1((\rho^1_s,\,\epsilon^{2,c}_s),\,(\epsilon^1_s,\,\epsilon^{2,d}_s))\,ds
\end{align*}
whence, taking the supremum over $\varphi\in\Lipone$ of both sides and using $t\leq T$ at the right-hand side,
$$ \bW_1((\rho^1_t,\,\epsilon^{2,c}_t),\,(\epsilon^1_t,\,\epsilon^{2,d}_t))\leq
	\cC' W_1(\rho^1_0,\,\epsilon^2_0)+\cC''\int_0^t\bW_1((\rho^1_s,\,\epsilon^{2,c}_s),\,(\epsilon^1_s,\,\epsilon^{2,d}_s))\,ds, $$
where we have set for brevity $\cC':=e^{\Lip{v^1}T}$ and $\cC'':=\sum_{\p=1}^{2}N^\p\Lip{v^\p}e^{\Lip{v^\p}T}$.

Applying Gronwall's inequality yields
$$ \bW_1((\rho^1_t,\,\epsilon^{2,c}_t),\,(\epsilon^1_t,\,\epsilon^{2,d}_t))\leq \cC'e^{\cC''t}W_1(\rho^1_0,\,\epsilon^2_0), $$
whence estimate~\eqref{eq:stability.est} follows straightforwardly with $\cC=\cC'e^{\cC''T}$.
\end{proof}

According to the proof of Theorem~\ref{theo:estimate}, an admissible constant $\cC$ in estimate~\eqref{eq:stability.est} is given explicitly by:
\begin{equation}
	\cC=\exp\left(\Lip{v^1}T+\sum_{\p=1}^{2}N^\p\Lip{v^\p}T\exp\left(\Lip{v^\p}T\right)\right).
	\label{eq:C}
\end{equation}
The constants $\Lip{v^\p}$, $\p=1,\,2$, appearing in this expression can be estimated taking into account, on the one hand, Proposition~\ref{prop:reg_v} and, on the other hand, the specific form of the interaction kernels~\eqref{eq:kernel} along with the biological considerations mentioned at the end of Section~\ref{sec:two_pop}. We have:
$$ \Lip{H^{\p\q}}=\max\left\{\frac{2F_r}{R_r},\,\frac{4F^{\p\q}_a}{R^{\p\q}_a-R_r}\right\} $$
whence
$$
	\begin{cases}
		\Lip{v^1}\leq N^1\max\left\{\dfrac{2F_r}{R_r},\,\dfrac{4F^{11}_a}{R^{11}_a-R_r}\right\}+N^2\max\left\{\dfrac{F_r}{R_r},\,\dfrac{4F^{12}_a}{R^{12}_a-R_r}\right\}, \\[5mm]
		\Lip{v^2}\leq k_0\Lip{\nabla{c}}+N^1\max\left\{\dfrac{2F_r}{R_r},\,\dfrac{4F^{21}_a}{R^{21}_a-R_r}\right\}+
			N^2\max\left\{\dfrac{2F_r}{R_r},\,\dfrac{4F^{22}_a}{R^{22}_a-R_r}\right\}.
	\end{cases}
$$
under the assumption $N^\p\geq 1$, $\p=1,\,2$. In particular, letting $R^{\p\q}_a=3R_r$ like in in Section~\ref{sec:simulations} gives $4F^{\p\q}_a/(R^{\p\q}_a-R_r)=2F^{\p\q}_a/R_r$. In this case, the above estimates for $\Lip{v^\p}$, $\p=1,\,2$, do not depend explicitly on the attraction radii $R^{\p\q}_a$.

\bigskip

In~\eqref{eq:stability.est} it is interesting to estimate directly also the term $W_1(\rho^1_0,\,\epsilon^1_0)$, so as to understand for which types of initial configurations of population $1$ one can expect the hybrid and the fully discrete models to be similar. The problem in its full generality is quite complex, therefore we will confine ourselves to a particular class of initial conditions coherent with the simulations presented in Section~\ref{sec:simulations}.

\begin{figure}[!t]
\centering
\includegraphics[width=0.4\textwidth]{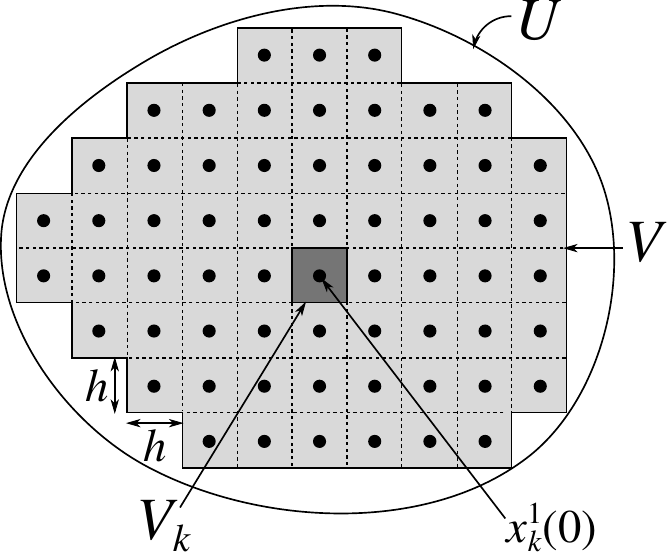}
\caption{Two-dimensional exemplification of the geometrical setting used to estimate $W_1(\rho^1_0,\,\epsilon^1_0)$.}
\label{fig.contdiscr}
\end{figure}

In more detail, we consider a constant density $\rho^1_0$ distributed on a bounded set $U\subset\Rn$ carrying the mass $N^1$, i.e.:
$$ \rho^1_0(x)=\frac{N^1}{\Leb{n}(U)}\chi_U(x), $$
where $\chi_U$ is the characteristic function of $U$. Parallelly we take $\epsilon^1_0=\sum_{k=1}^{N^1}\delta_{x^1_k(0)}$, where the $x^1_k(0)$'s are equispaced atoms within $U$, which we assume to be the centers of $n$-dimensional pairwise disjoint hypercubes $\{V_k\}_{k=1}^{N^1}$ of side length $h>0$ such that $\cup_{k=1}^{N^1}V_k=:V\subseteq U$, see Fig.~\ref{fig.contdiscr}.

\begin{theorem}[Estimate for $W_1(\rho^1_0,\,\epsilon^1_0$)]
Let $\rho^1_0,\,\epsilon^1_0\in\cM_1^{N^1}(\Rn)$ be as described above. Then:
$$ W_1(\rho^1_0,\,\epsilon^1_0)\leq N^1\left(\frac{\sqrt{n}}{2}+\frac{\Leb{n}(U\setminus V)}{\Leb{n}(U)}\right)\diam{U}, $$
where $\diam{U}=\ds{\sup_{x_1,\,x_2\in U}}\abs{x_2-x_1}$ is the diameter of $U$.
\label{theo:contdiscr}
\end{theorem}
\begin{proof}
Using the definition, we compute $W_1(\rho^1_0,\,\epsilon^1_0)$ as:
\begin{align*}
	W_1(\rho^1_0,\,\epsilon^1_0) &= \sup_{\varphi\in\Lipone}\int_{\Rn}\varphi\,d(\epsilon^1_0-\rho^1_0) \\
	&= \sup_{\varphi\in\Lipone}\left[\sum_{k=1}^{N^1}\varphi(x^1_k(0))-\frac{N^1}{\Leb{n}(U)}\int_{U}\varphi(x)\,dx\right];
\intertext{next we write $U=V\cup(U\setminus V)$ and use $V=\cup_{k=1}^{N^1}V_k$ (the union being disjoint by construction) and $\Leb{n}(V_k)=h^n$ to obtain:}
	&= N^1\sup_{\varphi\in\Lipone}\left[\sum_{k=1}^{N^1}\int_{V_k}\left(\frac{\varphi(x^1_k(0))}{N^1h^n}-\frac{\varphi(x)}{\Leb{n}(U)}\right)\,dx
		-\frac{1}{\Leb{n}(U)}\int_{U\setminus V}\varphi(x)\,dx\right],
\intertext{that is, noticing that $N^1h^n=\Leb{n}(V)$ and adding and subtracting $\varphi(x)/\Leb{n}(V)$ in the first integral,}
	&= N^1\sup_{\varphi\in\Lipone}\Biggl[\frac{1}{\Leb{n}(V)}\sum_{k=1}^{N^1}\int_{V_k}\bigl(\varphi(x^1_k(0))-\varphi(x)\bigr)\,dx\\
	&\phantom{=} +\left(\frac{1}{\Leb{n}(V)}-\frac{1}{\Leb{n}(U)}\right)\int_{V}\varphi(x)\,dx
		-\frac{1}{\Leb{n}(U)}\int_{U\setminus V}\varphi(x)\,dx\Biggr] \\
	&\leq N^1\sup_{\varphi\in\Lipone}\Biggl[\frac{1}{\Leb{n}(V)}\sum_{k=1}^{N^1}\int_{V_k}\abs{x^1_k(0)-x}\,dx \\
	&\phantom{=} +\frac{\Leb{n}(U\setminus V)}{\Leb{n}(U)}\left(\frac{1}{\Leb{n}(V)}\int_{V}\varphi(x)\,dx
		-\frac{1}{\Leb{n}(U\setminus V)}\int_{U\setminus V}\varphi(x)\,dx\right)\Biggr],
\end{align*}
where in the first integral we have used that $\varphi$ is Lipschitz continuous with $\Lip{\varphi}\leq 1$.

Since, by construction, $x^1_k(0)$ is the center of $V_k$ it results
$$ \abs{x^1_k(0)-x}\leq\frac{1}{2}\diam{V_k}=\frac{\sqrt{n}}{2}h \quad \forall\,x\in V_k. $$
Moreover, owing to the mean value theorem there exist $\bar{x}\in V$, $\tilde{x}\in U\setminus V$ such that
$$ \frac{1}{\Leb{n}(V)}\int_{V}\varphi(x)\,dx-\frac{1}{\Leb{n}(U\setminus V)}\int_{U\setminus V}\varphi(x)\,dx=
	\varphi(\bar{x})-\varphi(\tilde{x})\leq\abs{\bar{x}-\tilde{x}}\leq\diam{U}, $$
therefore
$$ W_1(\rho^1_0,\,\epsilon^1_0)\leq N^1\left(\frac{\sqrt{n}}{2}h+\frac{\Leb{n}(U\setminus V)}{\Leb{n}(U)}\diam{U}\right) $$
and the thesis follows by further using $h\leq\diam{U}$.
\end{proof}

\begin{remark}
Since $\frac{\Leb{n}(U\setminus V)}{\Leb{n}(U)}\leq 1$, a rougher estimate which does not involve $V$ is
$$ W_1(\rho^1_0,\,\epsilon^1_0)\leq N^1\left(\frac{\sqrt{n}}{2}+1\right)\diam{U}. $$
\end{remark}

\section{Discussion}
\label{sec:discussion}
Biological systems are intrinsically multiscale, hence complex. The collective evolution of cell aggregates is in fact determined by the behavior and the biophysical properties of single cells, as well as by their mutual interactions. A fundamental feature of mathematical approaches for biological/biomedical problems is therefore the ability to describe and reproduce such a multilevel framework. Motivated by this argument, in this paper we have introduced a mathematical method, based on measure theory, intended to provide an ensemble representation of a cell population starting from an individual-based characterization of the phenomenology of a representative component cell. In more detail, the key idea of the proposed approach is to describe the global evolution of a cell aggregate in terms of a time-evolving mass measure properly abstracted from the individual-based law accounting for single cell dynamics. This way the scale of representation of the underlying physical system, i.e., microscopic/discrete vs. macroscopic/continuous, can be chosen \textit{a posteriori} according only to the spatial structure given to the mass measure.  In our opinion, the proposed method offers several conceptual advantages in modeling biological phenomena: first, it enables one to unambiguously transfer cell-level concepts, such as the cell interaction radii and forces, into a super-cellular description of cell populations. Second, it is particularly suited to represent, within a unified modeling framework, biological systems characterized by the coexistence of different functional subsystems, each of which needs a proper spatial description linked to the specific biophysical determinants. In these cases, it is in fact necessary to treat each cell phenotype at a distinct scale according to its intrinsic properties. In this respect, we have focused the main part of the paper on the analysis of a biological system composed of an inactivated/undifferentiated cell colony, properly described with a continuous approximation, and of a small number of activated/differentiated cells, singularly represented as discrete point masses. The resulting realizations have shown that the evolution of the overall system is mainly driven by the behavior of the activated cells, which guide the inactivated population. Specifically, the emerging patterns arise from the strength of the exogenous interactions, as its value determines whether the activated individuals are free to scatter in the neighboring space (eventually followed by different types of finger structures of the undifferentiated population) or are absorbed by the continuous mass itself.

\begin{figure}[!t]
\centering
\includegraphics[width=1\textwidth]{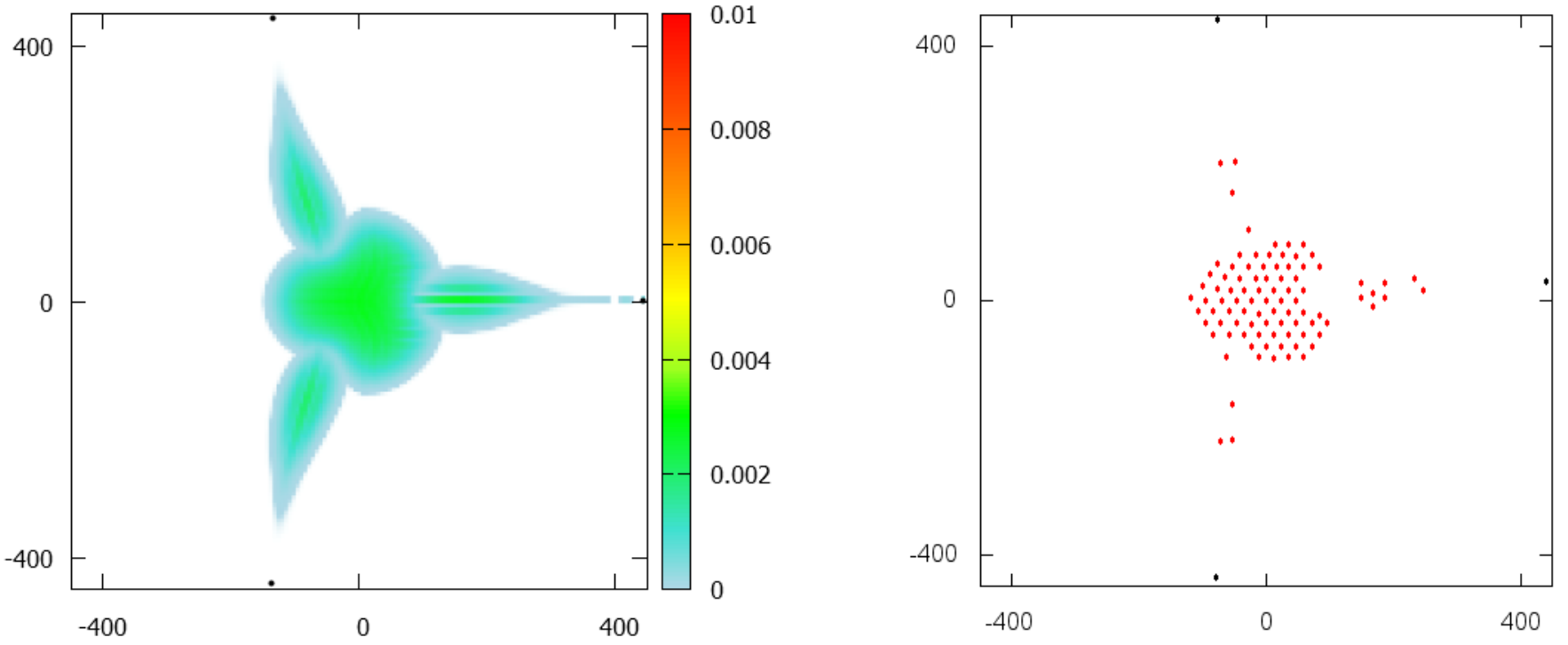}
\caption{Final configurations (\ie, at $T=3600\Delta{t}\approx 12\unit{h}$) of the two-population system, where population 1 is represented either with a continuous approximation or with a discrete description. In both cases, the parameter setting is the same as that of the simulations presented in Fig.~\ref{fig:different_N2}.}
\label{fig:hybrid_discr_discuss}
\end{figure}

In all simulations proposed in Section~\ref{sec:simulations}, the colony of undifferentiated cells has been represented with a continuous approximation. However, one may inquire into the differences in system dynamics arising from switching to a conceivable discrete description. In this respect, we have devoted Section~\ref{sec:comparison} to some analytical results, which deal with this interesting multiscale aspect. In particular, the resulting stability estimates of the discrepancy in the evolution between the hybrid and fully discrete solutions of the two-population system depend on both the initial condition and selected biologically relevant quantities. In more detail, from~\eqref{eq:C} we see that the smaller $\Lip{v^\p}$ the smaller the constant $\cC$, which, owing to Theorem~\ref{theo:estimate}, is a sort of amplification factor of the initial discrepancy between the continuous and discrete descriptions of population $1$. Hence, for fixed initial conditions $\rho^1_0$, $\epsilon^1_0$, population size $N^\p$, and final time $T$, we can state that decrements in the strength of the intercellular interactions $F_r$, $F^{\p\q}_a$, as well as increments in the corresponding radii $R_r$, $R^{\p\q}_a$, have a stabilizing effect on the evolution of the discrepancy at successive times. Furthermore, according to Theorem~\ref{theo:contdiscr}, if the same mass $N^1$ is initially homogeneously distributed on a smaller and smaller set $U$ then the similarity between the continuous and discrete descriptions at the initial time $t=0$, hence also at successive times $t>0$ in view of the previous theorem, increases. If, for instance, $U$ is a hypercube in $\Rn$ and if $N^1$ is such that $\sqrt[n]{N^1}\in\mathbb{N}$ then one can take $V=U$, whence $W_1(\rho^1_0,\,\epsilon^1_0)\leq \frac{1}{2}N^1\sqrt{n}\diam{U}=\frac{1}{2}N^1nl_0$, where $l_0>0$ is the edge length of $U$. In two dimensions ($n=2$) this becomes $W_1(\rho^1_0,\,\epsilon^1_0)\leq N^1l_0$, which scales linearly with the characteristic length $l_0$. If instead, referring to the specific simulations presented in Section~\ref{sec:simulations}, $U$ is a circle in $\R^2$ with radius $r_0>0$ then the exact computation of $\Leb{2}(U\setminus V)$ is harder. However, since $\diam{U}=2r_0$, from Theorem~\ref{theo:contdiscr} it is possible to conclude that $W_1(\rho^1_0,\,\epsilon^1_0)$ scales again at least linearly with the characteristic length $r_0$. Figure~\ref{fig:hybrid_discr_discuss} shows that, under the assumptions of Theorems~\ref{theo:estimate},~\ref{theo:contdiscr}, the evolutions of the hybrid and fully discrete systems are closely comparable. Notice in particular the formation of fingers and the shape of the central core of population $1$.

\paragraph{Comparison with the pertinent literature}
The hybrid coupling of continuous-discrete descriptions, here implemented in the two-population cell system, is an increasingly common approach for biological/biomedical problems. First, a wide range of modeling frameworks use a discrete representation for cell-scale elements and a continuous approximation to describe the evolution of subcellular molecular variables (e.g., nutrients, oxygen, and diffusible factors) that may evolve both within the cells and in the extracellular environment. This is typically done in most cellular Potts models~\cite{Scianna2012}. In this respect, in~\cite{Anderson2006} a multilevel approach studying tumor morphology and phenotypic evolution under different pressure conditions treats malignant cells as discrete stochastic variables and microenvironmental chemicals as deterministic concentrations. As a result, the authors point out that invasive masses, characterized by fingering margins and dominated by few clones with aggressive traits, are observed in hypoxic conditions. Similarly, in~\cite{Drasdo2009} an individual cell-based model describes a cancer mass and the surrounding vasculature (both preexisting and tumor-derived) and accounts for contact-inhibition of growth and nutrient-related growth and apoptosis, while the kinetics of oxygen and of short- and long-range diffusing angiogenic factors are modeled with standard reaction-diffusion (RD) equations. Further, in~\cite{RamisConde2008} the authors approach cancer cell invasion with a multiscale individual-based lattice free model, where each tumor cell is individually represented as an isotropic elastic object capable of division and evolving following a master equation, while intracellular dynamics of the E-cadherin-$\beta$-catenin interactions are treated with a system of RD equations. In~\cite{Kim2013} the authors present a model for the study of the evolution of a ductal carcinoma in situ (an early non-invasive stage of breast cancer) in which epithelial cells are modeled individually while the extracellular matrix (ECM) forming the tumor microenvironment is treated as a continuum. Finally, in the very recent work~\cite{DiCostanzo2014} the authors propose a hybrid model for the morphogenesis of the posterior lateral line system in zebrafishes, which uses a discrete description for the cellular level (comprising both differentiated and undifferentiated cells) and a continuous one for the molecular level (chemical signals).

Other families of hybrid models developed to study the evolution of tumor spheroids use instead different mathematical descriptions for malignant cells located in different regions of the malignant mass. These approaches are conceptually more similar to the one proposed in this work. For instance, in~\cite{Kim2007,Stolarska2009} an agent-based model (ABM) is implemented to describe the proliferating and highly metabolic malignant cells at the spheroid edge (specifically, their interactions, morphology, and internal dynamics), while a continuous viscoelastic model is used for the quiescent and necrotic regions within the bulk of the tumor mass. Analogously, in~\cite{Tanaka2009} an ABM describes the invading cells of glioma tumors as discrete dimensionless points, whereas the rest of the malignant mass is approached with a continuous model suitable to maintain a spherical symmetry. In~\cite{Frieboes2010} a further ABM is used to represent differentiated mesenchymal cells and is coupled to a continuous model for undifferentiated epithelial volume fractions. In particular, discrete cells are set to be produced in hypoxic regions: they can be locally converted back to the continuous description if their number exceeds a given threshold (this is the so-called \textit{epithelial-to-mesenchymal transition}). Finally, in~\cite{Chauviere2012} the authors propose a modeling approach for tumor growth based on the Dynamic Density Functional Theory, which is able to provide a mesoscale continuous framework directly derived from a stochastic discrete model. Such a mathematical framework can be extended in order to take into account different cell subpopulations characterized by different phenotypes and birth/death processes. Of particular interest is also the idea proposed in~\cite{Anderson1998,Chaplain2000} to derive a discrete stochastic model from a continuous deterministic model for reproducing different aspects of tumor-induced angiogenesis. In more detail, the continuous approach consists of a system of nonlinear partial differential equations describing the chemotactic and haptotactic migration of endothelial cells, while the corresponding discrete model is based on a discretized form (via a finite difference approximation) of the system of PDEs. This way, on one side the parameter values used in the discrete model are directly related to those of the continuous model; on the other side, phenomena like branching and anastomosis, and their dependence on subcellular mechanisms, can be properly incorporated.

\paragraph{Model refinements}
As commented in the first part of this conclusive section, we expect our method to give an interesting contribution in the development of multiscale approaches to the modeling of cell phenomena. However, the presented mathematical framework has still some important limitations, whose overcoming would give rise to relevant modeling improvements. For instance, it would be interesting to describe in more detail cell interactions with the extracellular environment, for instance by including haptotactic or durotactic mechanisms or the possible proteolysis of ECM insoluble fibrous components, which is a critical determinant of cell migration especially in three-dimensional highly constrained environments~\cite{Scianna2013,Wolf2007}.

However, the main open issue is the fact that, in the present version of the approach, cell duplication/death, as well as cell differentiation in the case of multiple coexisting populations, are not included. This is, of course, a relevant point for a detailed simulation of a wide spectrum of biological phenomena (such as e.g., morphogenesis, tumor growth, wound healing). Taking into account such processes requires significant revisions of the analytic architecture underlying the model. In particular, the mass measures no longer evolve satisfying conservation equations, since localized source/sink terms have to be included (see~\cite{Piccoli2014} for theoretical insights). Even more important, as far as the derivation of the model is concerned, the passage from the phenomenological law of test individuals to the collective model of the corresponding cell population can no longer rely only on the use of conservation and transport theorems as proposed in Section~\ref{sec:formal_derivation}. In fact the hybrid evolution equation \eqref{eq:test_cell} for the representative cell $X^\p_t$ needs, on the one hand, to account for the specific localization of mitotic/apoptotic processes, which is not trivial, and, on the other hand, to be coupled with a law for the time variation of the number $N^\p(t)$ of individuals of the population.

Challenging questions result also from cell phenotypic transitions, i.e., mass exchanges among different populations, for example between a continuous aggregate and a discrete set of differentiated cells. In particular, if the replacement of a discrete cell element with a corresponding continuous approximation does not represent an issue (provided the mass is conserved), the opposite transition is much more complicated, as some discontinuities, density holes, or even subdomains with negative densities may arise, thereby creating analytical and numerical criticalities.

\section*{Acknowledgments}
The authors extend warm thanks to Luigi Preziosi for many stimulating and fruitful discussions.

\bibliographystyle{plain}
\bibliography{CaSmTa_multiscale_cell}

\end{document}